\documentclass[twocolumn,astrosymb,tighten,times]{aastex631}
\usepackage{mathtools}
\usepackage{xspace}
\usepackage{verbatim}

\usepackage{ulem}

\received{September 26, 2021}
\revised{May 6, 2022; March 3, 2023}
\accepted{March 9, 2023}
\submitjournal{\apj}

\shorttitle{Heavily X-ray Obscured AGNs}
\shortauthors{Carroll et al.}
\graphicspath{{./}{./}}

\begin{document}

\title{A High Fraction of Heavily X-ray Obscured Active Galactic Nuclei}

\author[0000-0003-3574-2963]{Christopher M. Carroll}
\affiliation{Department of Physics and Astronomy, Washington State University, 1245 Webster Hall, Pullman, WA 99164, USA}
\affiliation{Department of Physics and Astronomy, Dartmouth College, 6127 Wilder Laboratory, Hanover, NH 03755, USA}

\author[0000-0001-8211-3807]{Tonima T. Ananna}
\affiliation{Department of Physics and Astronomy, Dartmouth College, 6127 Wilder Laboratory, Hanover, NH 03755, USA}

\author[0000-0003-1468-9526]{Ryan C. Hickox}
\affiliation{Department of Physics and Astronomy, Dartmouth College, 6127 Wilder Laboratory, Hanover, NH 03755, USA}

\author[0000-0002-7100-9366]{Alberto Masini}
\affiliation{SISSA, Via Bonomea 265, 34151 Trieste, Italy}
\affiliation{INAF -- Osservatorio di Astrofisica e Scienza dello Spazio di Bologna, via Gobetti 93/3, I-40129 Bologna, Italy}

\author[0000-0002-9508-3667]{Roberto J. Assef}
\affiliation{N\'{u}cleo de Astronom\'{i}a de la Facultad de Ingenier\'{i}a y Ciencias, Universidad Diego Portales, Av. Ej\'{e}rcito Libertador 441, Santiago, Chile}

\author[0000-0003-2686-9241]{Daniel Stern}
\affiliation{Jet Propulsion Laboratory, California Institute of Technology, 4800 Oak Grove Drive, Mail Stop 169-221, Pasadena, CA 91109, USA}

\author[0000-0002-4945-5079]{Chien-Ting J. Chen}
\affiliation{Astrophysics Office, NASA Marshall Space Flight Center, ZP12, Huntsville, AL 35812, USA}

\author[0000-0002-3249-8224]{Lauranne Lanz}
\affiliation{Department of Physics, The College of New Jersey, 2000 Pennington Road, Ewing, NJ 08628, USA}

%%%%%%%%%%%%%%%%%%%%%%%%%%%%%%%%%%%%%%%%%%%%%%%%%%%%%%%%%%%
%%
%%      Abstract
%%
%%%%%%%%%%%%%%%%%%%%%%%%%%%%%%%%%%%%%%%%%%%%%%%%%%%%%%%%%%%
\begin{abstract}
We present new estimates on the fraction of heavily X-ray obscured, Compton-thick (CT) active galactic nuclei (AGNs) out to a redshift of $z \leq$ 0.8. From a sample of 540 AGNs selected by mid-IR (MIR) properties in observed X-ray survey fields, we forward model the observed-to-intrinsic X-ray luminosity ratio ($R_{L_{\text{X}}}$) with a Markov chain Monte Carlo (MCMC) simulation to estimate the total fraction of CT AGNs ($f_{\text{CT}}$), many of which are missed in typical X-ray observations. We create model $N_{\text{H}}$ distributions and convert these to $R_{L_{\text{X}}}$ using a set of X-ray spectral models. We probe the posterior distribution of our models to infer the population of X-ray non-detected sources. From our simulation we estimate a CT fraction of \mbox{$f_{\text{CT}}$ = $\text{0.555}^{+\text{0.037}}_{-\text{0.032}}$}. We perform an X-ray stacking analysis for sources in Chandra X-ray Observatory fields and find that the expected soft (0.5--2 keV) and hard (2--7 keV) observed fluxes drawn from our model to be within 0.48 and 0.12 dex of our stacked fluxes, respectively. Our results suggests at least 50\% of all MIR-selected AGNs, possibly more, are Compton-thick \mbox{($N_{\text{H}} \gtrsim$ 10$^{\text{24}}$ cm$^{-\text{2}}$)}, which is in excellent agreement with other recent work using independent methods. This work indicates that the total number of AGNs is higher than can be identified using X-ray observations alone, highlighting the importance of a multiwavelength approach. A high $f_{\text{CT}}$ also has implications for black hole (BH) accretion physics and supports models of BH and galaxy co-evolution that include periods of heavy obscuration.
\end{abstract}


%%%%%%%%%%%%%%%%%%%%%%%%%%%%%%%%%%%%%%%%%%%%%%%%%%%%%%%%%%%
%%
%%      Introduction
%%
%%%%%%%%%%%%%%%%%%%%%%%%%%%%%%%%%%%%%%%%%%%%%%%%%%%%%%%%%%%
\section{Introduction} \label{sec:intro}
Supermassive black holes (SMBHs) are found ubiquitously at the centers of massive galaxies, growing via intermittent accretion of interstellar material and occasional SMBH mergers. During intense periods of gas accretion, these active galactic nuclei (AGNs) are extremely luminous, radiating across the electromagnetic spectrum. Since the seminal discovery that obscuring dust is responsible for the observed difference between broad-line and narrow-line AGNs (\citealp{antonucci1985}; see \citealp{ricci2017b,ananna2022a,ananna2022b} for more recent models of unification), major efforts have been undertaken to characterize the nature and extent of the obscuring material, with attempts to determine its composition \citep[e.g.,][]{davies2015,garcia-burillo2016} and physical scale \citep[e.g.,][]{honig2012,tristram2014,chen2015,lopez-gonzaga2016,panagiotou2019}. Obscuration across electromagnetic regimes is also thought to originate from different sources, namely nuclear gas producing absorption at X-ray energies and circumnuclear dust grains causing extinction of UV and optical emission \citep[see][]{hickox2018}.

Infrared surveys have provided evidence that more than half of the AGN population exhibit signatures of obscuration \citep{lacy2013}. Optical--near-IR (NIR) colors have proven reliable in selecting obscured sources in large AGN samples \citep[e.g.,][]{hickox2007b,lamassa2016} but lack the power to accurately estimate levels of obscuration, particularly in heavily obscured sources where the host galaxy dominates at optical and NIR wavelengths. Mid-IR (MIR) techniques have further improved our ability to probe heavily obscured AGN that are missed by optical surveys \citep[e.g.,][]{stern2005,stern2012,mateos2013,assef2015}, as MIR observations probe emission from dust and are relatively unaffected by dust obscuration.

While MIR observations have contributed reliable techniques for identifying obscured AGNs, X-ray spectroscopic observations have become the dominant method of probing nuclear obscuring gas content \citep[see][]{hickox2018}, the bulk of which need not necessarily trace the distribution of dust. Though the X-ray regime has become the favored method for measuring obscuring column densities ($N_{\text{H}}$), recent work has shown that hard X-ray energies ($\geq$2--10 keV) can remain undetected for heavily obscured AGN in typical observations \citep[e.g.,][]{lansbury2015,yan2019,lambrides2020}. This provides a particular challenge for Compton-thick (CT) densities (i.e., \mbox{$N_{\text{H}} \geq$ 1.5$\times$10$^{\text{24}}$ cm$^{-\text{2}}$}), where scattering optical depths reach order unity \citep[e.g.,][]{lansbury2015}. As deep, hard X-ray observations are required to positively identify and constrain the density of obscuring material surrounding CT AGNs \citep[e.g.,][]{traina2021}, large-scale searches have been limited to smaller fields. To date there has been no reliable direct measure of the complete fraction of CT AGN ($f_{\text{CT}}$), marking a major uncertainty in the total number of accreting black holes (BHs) in the universe. New techniques are therefore required to accurately measure the line-of-sight gas densities for the most heavily obscured AGNs, and enable a complete view of the population \citep[e.g.,][]{pfeifle2021}.

One compelling approach involves utilizing the empirical relationship between the X-ray and MIR luminosities ($L_{\text{X}}$ and $L_{\text{MIR}}$ respectively) of AGNs \citep[i.e.,][]{fiore2009,stern2015,chen2017}. The combination of X-ray and MIR luminosities has been shown to provide a reliable method for identifying extremely obscured AGNs ($N_{\text{H}} >$ 10$^{\text{24}}$ cm$^{-\text{2}}$) in large samples \citep[e.g.,][hereafter C21]{carroll2021}. The \mbox{$L_{\text{X}}$--$L_{\text{MIR}}$} relation provides a method of inferring the intrinsic X-ray luminosity of an AGN through its MIR luminosity ($L_{\text{X}}(L_{\text{MIR}})$), which is reasonably straightforward to estimate given the quality and depth of all-sky IR data. The ratio of the observed-to-intrinsic X-ray luminosity ($R_{L_{\text{X}}}$) then provides insight into the amount of obscuring material present in a given source and can be calculated directly from observables. X-ray spectral models provide the final component, allowing us to understand the effects of increasing $N_{\text{H}}$ on observed X-ray fluxes and ultimately $R_{L_{\text{X}}}$.

Moreover, C21 showed that the same technique can be used to estimate lower limits on $N_{\text{H}}$ for sources that are not detected in X-ray observations, substituting X-ray flux limits in lieu of direct detections. However, the $R_{L_{\text{X}}}$ distribution for X-ray detected and non-detected sources differs drastically and cannot be directly compared, as $R_{L_{\text{X}}}$ for non-detected sources are upper limits. As such, $N_{\text{H}}$ can only be estimated with any given accuracy for X-ray detected sources using this methodology. Therefore, a new statistical method is required to include the X-ray non-detected sources in our estimates of the distribution of $N_{\text{H}}$.

Previous studies have estimated the space density of AGNs---including CT sources---predominantly through modeling of the X-ray luminosity function \citep[XLF; i.e.,][]{ueda2014,buchner2015,aird2015b,fotopoulou2016,ananna2019}, which describes the number density of AGNs as a function of X-ray luminosity and redshift. This approach is dependent on large samples of X-ray selected AGNs in soft and hard energies. Additionally, this method is generally reliant on assuming an absorption function to account for obscuration and/or modeling of obscured and unobscured sources separately. Multiple observational constraints may be used in the generation of the XLF through synthesis models, including X-ray number counts, spectral shape of the integrated cosmic X-ray background (CXB), and the observed fraction of CT AGNs. The latest synthesis models produce an XLF out to $z$ = 1 that accounts for all observational constraints \citep[][hereafter A19]{ananna2019}, predicting a CT fraction of 0.50$\pm$0.09 for low- to moderate-luminosity AGNs \mbox{(log $L_{\text{X}}/\text{erg s}^{-\text{1}} <$ 43.6)}, and as high as 0.56$\pm$0.07 at higher luminosities (log \mbox{$L_{\text{X}}/\text{erg s}^{-\text{1}} >$ 43.6})---much higher than previous studies.

In this work, we aim to make a direct measurement of $f_{\text{CT}}$ and determine the underlying $N_{\text{H}}$ distribution of heavily obscured sources \mbox{($N_{\text{H}} \geq$ 10$^{\text{24}}$ cm$^{-{\text{2}}}$)}. Our approach relies on multiwavelength archival data to account for AGNs which are either undetected or poorly detected in the X-rays. We use a novel multiwavelength, non-parametric approach to forward model the $N_{\text{H}}$ distribution by simulating the observed $R_{L_{\text{X}}}$ distribution in a complete sample of AGNs.

This paper is organized as follows: Section \ref{sec:data} details our photometric sample and matching X-ray fields; Section \ref{sec:modeling} describes our modeling procedure; Section \ref{sec:analysis} details our analysis and results of our modeling; Section \ref{sec:discussion} summarizes our results and future work. Throughout the paper, we assume a $\Lambda$CDM cosmology with parameters \mbox{$H_{\text{0}}=$ 70 km s$^{-{\text{1}}}$ Mpc$^{-{\text{1}}}$}, \mbox{$\Omega_{\text{m}}=$ 0.3}, and \mbox{$\Omega_\Lambda=$ 0.7} (\citealt{spergel2007}).

%%%%%%%%%%%%%%%%%%%%%%%%%%%%%%%%%%%%%%%%%%%%%%%%%%%%%%%%%%%
%%
%%      Data
%%
%%%%%%%%%%%%%%%%%%%%%%%%%%%%%%%%%%%%%%%%%%%%%%%%%%%%%%%%%%%
\section{Data} \label{sec:data}
The sample adopted in this paper builds on previous work by C21, having identified heavily obscured AGNs from multiwavelength archival data. The details of their data selection and modeling are fully described therein and briefly summarized here.

Our sample consists of 540 AGNs selected on the basis of MIR colors \citep{assef2018} from the Wide-field Infrared Survey Explorer \citep[WISE;][]{wright2010} and matched to Sloan Digital Sky Survey \citep[SDSS;][]{york2000} optical counterparts. Each source in our sample also lies within fields observed by the Chandra \citep{evans2020}, XMM-Newton \citep{rosen2016}, and/or NuSTAR \citep{civano2015,lansbury2017,masini2018a} X-ray observatories. Although the 540 sources in our final sample all lie within observed X-ray fields, only 40\% have X-ray detections, with the remaining 60\% lacking X-ray counterparts. This working sample was drawn from a larger parent dataset of over three million sources with redshift measurements (photometric and spectroscopic), each with matching SDSS--WISE photometry.

Optical photometry for our sample was obtained from a combination of SDSS Data Release 14 \citep{abolfathi2018} and XDQSO\textit{z} \citep{dipompeo2015}. In this work, we chose to replace all MIR photometry from the WISE AllWISE Source Catalog with unWISE \citep{lang2014}. The unWISE Catalog produced WISE forced photometry at known SDSS source positions, ensuring reliable source association across catalogs. Use of the unWISE Catalog increases the depth of WISE observations at wavelengths crucial to disentangling absorbed and non-absorbed systems, all while maintaining the highest possible signal-to-noise ratio (S/N). Additional photometry was added in UV and NIR bands where available from the Galaxy Evolution Explorer \citep[GALEX;][]{martin2005}, the UKIRT Infrared Deep Sky Survey \citep[UKIDSS;][]{lawrence2007}, and the Two Micron All Sky Survey \citep[2MASS;][]{skrutskie2006} (see C21 for additional details).

To ensure the accuracy of our modeling and minimize systematics, we imposed several quality selection criteria to our data. First, we required all available photometry to have a minimum S/N $\geq$ 3.0. We required a minimum of seven photometric bands per source for adequate coverage to model spectral energy distributions (SEDs), with the additional requirement of detections in all four WISE bands to accurately constrain AGN contribution to the MIR. SDSS photometry was only considered where the \mbox{PhotoObjAll} column \mbox{\texttt{clean} = 1}, ensuring reliable photometry. Though our data was primarily WISE selected, all sources are detected in $gri$. Finally, we applied an angular bright stars mask to remove regions of the sky with possible IR contamination \citep{dipompeo2014b}. We chose to limit sources to redshifts $z \leq$ 0.8 and only used SDSS photometric redshifts where the Photoz column \texttt{photoErrorClass} was {$-$}1, 1, 2, or 3, corresponding to an average root-mean-square error between \mbox{0.066 $\leq$ RMSE $\leq$ 0.074}. Our final sample consists of sources having 64\% photometric and 36\% spectroscopic redshifts.

\begin{figure*}
    \plottwo{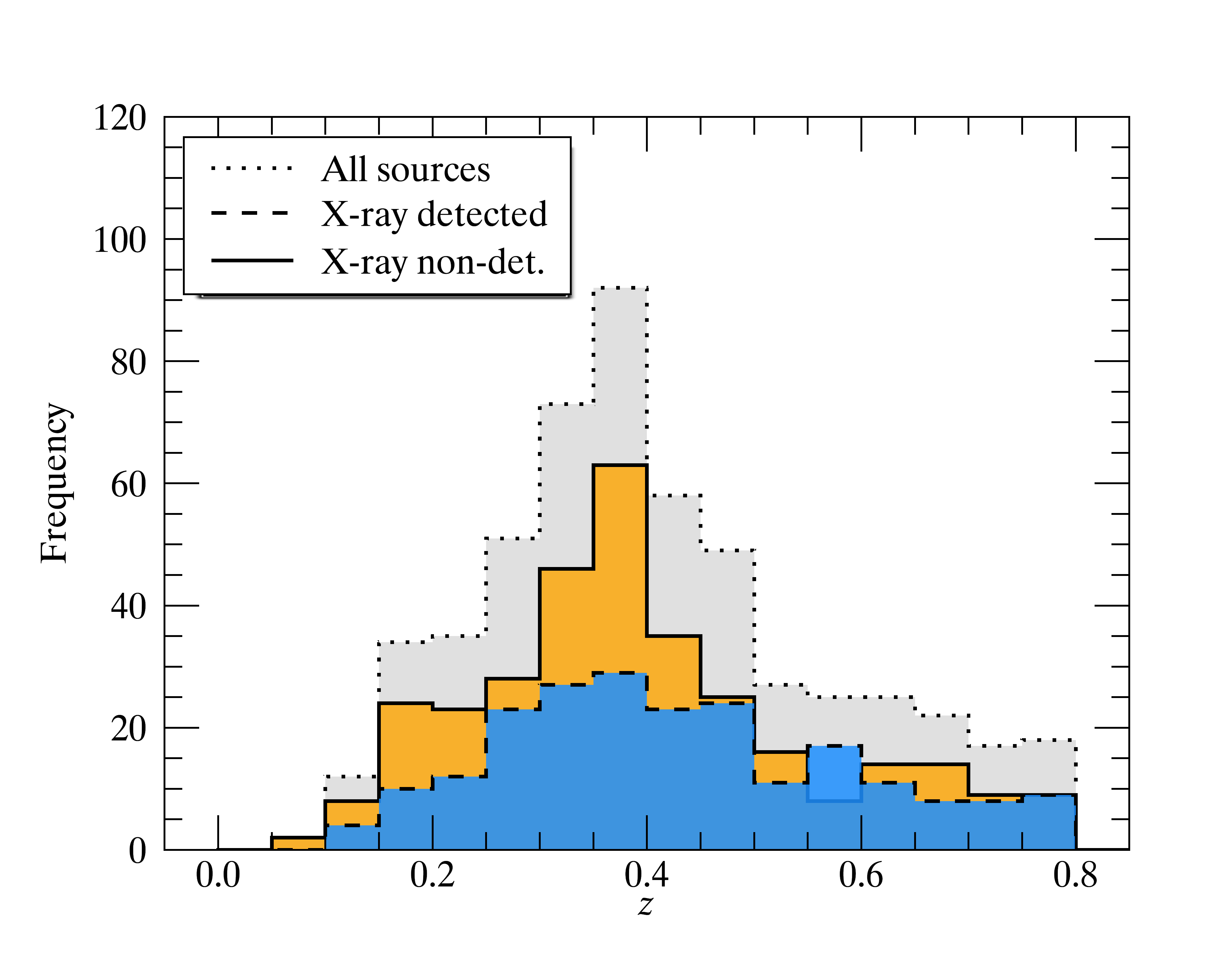}{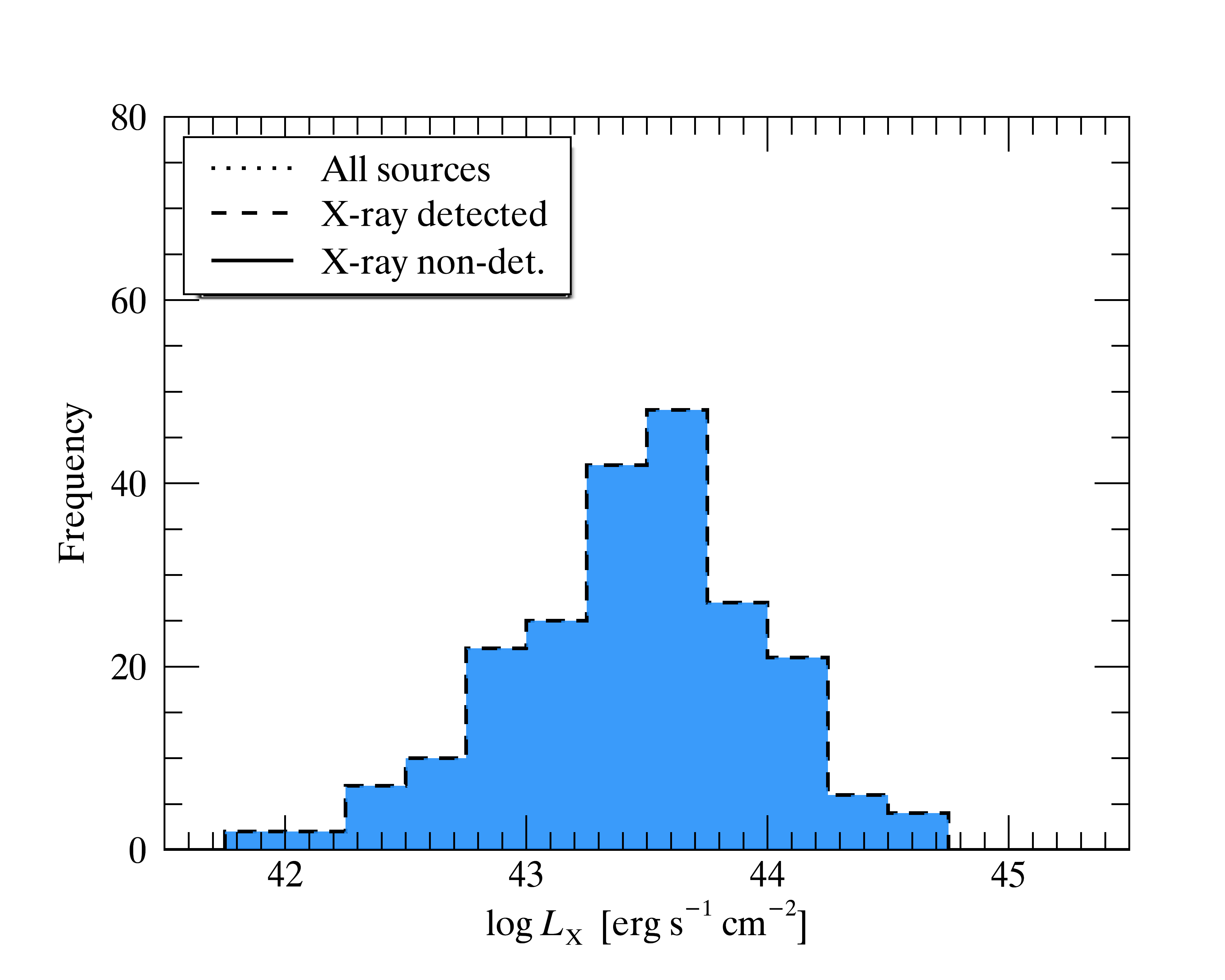}
    \plottwo{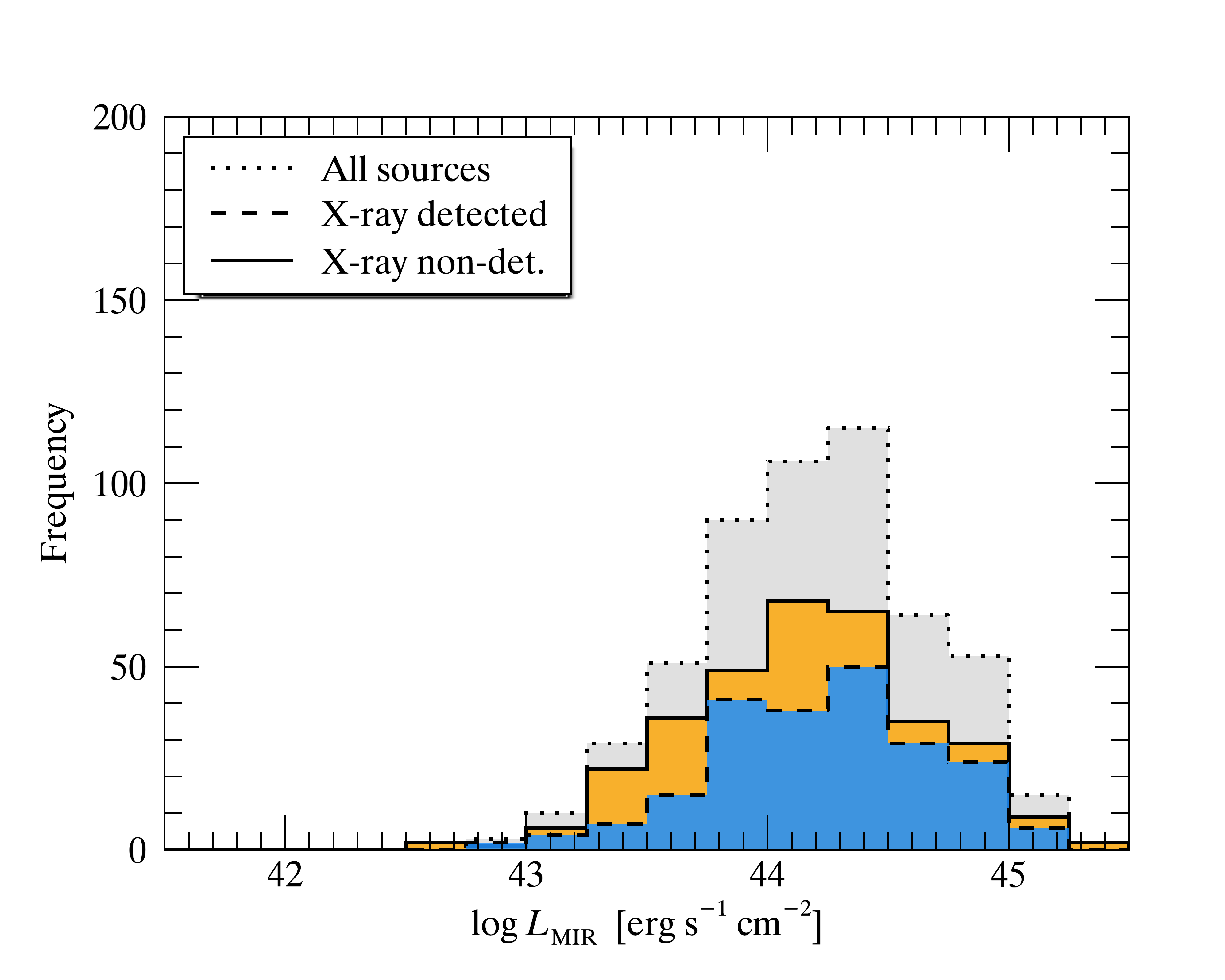}{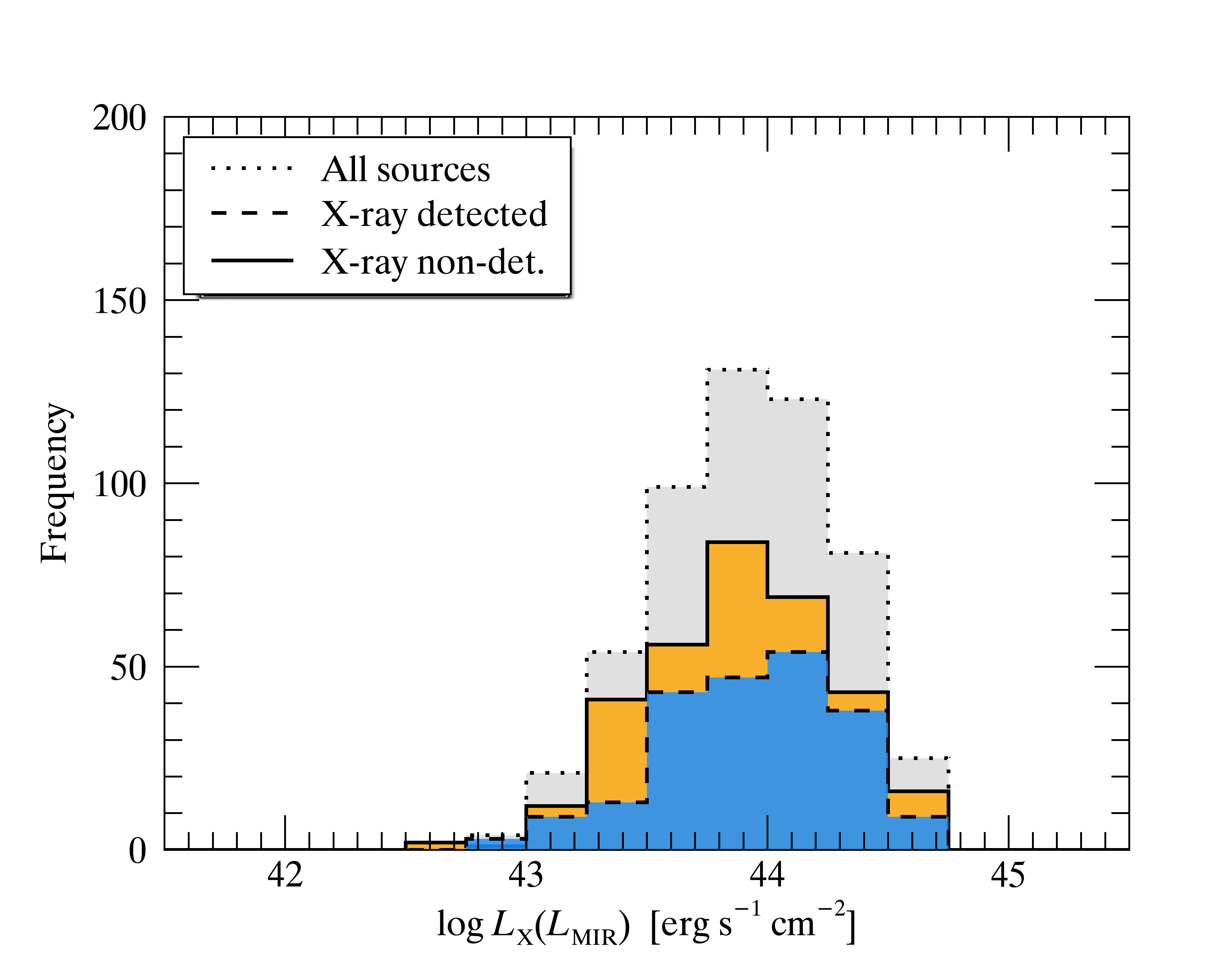}
    \caption{Distribution of sample properties: redshift ($z$; top left); observed 2--10 keV X-ray luminosity ($L_{\text{X}}$; top right); MIR AGN \mbox{6 $\mu$m} luminosity ($L_{\text{MIR}}$; bottom left); intrinsic 2--10 keV X-ray luminosity estimates ($L_{\text{X}}(L_{\text{MIR}})$; bottom right). In each panel, the distributions are demarcated as X-ray detected (dashed blue) and X-ray non-detected (solid orange) sources, along with the comprehensive distribution (dotted gray).}
    \label{fig:properties}
\end{figure*}

Broadband SEDs were modeled using a combination of galaxy and AGN templates \citep{assef2010,kirkpatrick2015}. From our SED modeling, sources were only considered if they passed a minimum reduced chi-square cut ($\chi^{\text{2}}_{\text{red}} \leq$ 20.0) to reject poor fits. We also required sources to derive a high fraction of their total MIR SED emission from AGN contribution ($f_{\text{AGN}} \geq$ 0.7 at 15$\mu$m), and further limited our final sample to IR-luminous sources \mbox{($L_{\text{MIR}} >$ 10$^{\text{42}}$ erg s$^{-\text{1}}$)}.

All sources in our final sample contain MIR 6 $\mu$m AGN luminosities ($L_{\text{MIR}}$) drawn directly from our SED modeling. Each object lies within an X-ray observed field and includes 3$\sigma$ upper-limit X-ray luminosities ($L_{\text{X-lim}}$) for both X-ray detected and non-detected sources, drawn from the respective flux limits of each instrument (see Figure 3 of C21). Observed 2--10 keV X-ray luminosities ($L_{\text{X}}$) were calculated for the X-ray detected sources in our sample. The observed $L_{\text{X}}$--$L_{\text{MIR}}$ relation of \cite{chen2017} was used to estimate intrinsic X-ray luminosities ($L_{\text{X}}$($L_{\text{MIR}}$)) for all sources, corresponding to their derived MIR AGN luminosities. A breakdown of the sample properties described in this section are shown in Figure \ref{fig:properties} and are separated for comparison into X-ray detected and non-detected sources. Lastly, we defined the ratio of observed-to-intrinsic X-ray luminosity as
\begin{equation}\label{eq:rlx}
    R_{L_{\text{X}}} = 
    \begin{dcases*}
    \frac{L_\text{X}}{L_{\text{X}}(L_{\text{MIR}})} & \quad for X-ray detected, \\
    \frac{L_{\text{X-lim}}}{L_{\text{X}}(L_{\text{MIR}})} & \quad for X-ray non-detected,
    \end{dcases*}
\end{equation}
which is used as a proxy for nuclear obscuration.

%%%%%%%%%%%%%%%%%%%%%%%%%%%%%%%%%%%%%%%%%%%%%%%%%%%%%%%%%%%
%%
%%      Modeling
%%
%%%%%%%%%%%%%%%%%%%%%%%%%%%%%%%%%%%%%%%%%%%%%%%%%%%%%%%%%%%
\section{Modeling} \label{sec:modeling}

%%%%%%%%%%%%%%%%%%%%%%%%%%%%%%%%%%%%%%%%%%%%%%%%%%%%%%%%%%%
%%      X-ray Spectral Models
%%%%%%%%%%%%%%%%%%%%%%%%%%%%%%%%%%%%%%%%%%%%%%%%%%%%%%%%%%%
\subsection{X-Ray Spectral Models} \label{ssec:xspec}
We used a set of X-ray spectral models to estimate the impact of nuclear obscuration on observed AGN X-ray flux. Our spectral models were constructed using the X-ray spectral fitting package \textsc{xspec} \citep{arnaud1996} with the following syntax:
\begin{multline*}
\textsc{constant}\times\textsc{cutoffpl}\\
+\,\textsc{zphabs} \times \textsc{cabs} \times \textsc{pexmon}\\
+\,\textsc{borus02}.
\end{multline*}
Each of our models follow a similar prescription, 
being the sum of transmitted, reflected, and scattered components to account for the diversity of AGN X-ray spectra \citep[e.g.,][]{buchner2014,ananna2020a}.

We began with an initial spectral model \textsc{cutoffpl}, consisting of a 300 keV high-energy cutoff power law to represent the intrinsic continuum of UV--optical photons upscattered to X-ray energies through Comptonization by the AGN corona. To account for Thomson scattering of the X-ray continuum by photoionized, circumnuclear material, we assumed the variable scattering fraction of \citet[][hereafter, G21]{gupta2021}, applied as a multiplicative constant to the cutoff power law. G21 observe that the scattering fraction ($f_{\text{scatt}}$) decreases with increasing $N_{\text{H}}$ and shows no dependency on X-ray luminosity or BH mass. To encapsulate any other potential factors due to AGN diversity in our sample, we chose to incorporate the G21 uncertainties on scattering fraction ($\sigma_{\text{scatt}}$) as a parameter in our model. Though we technically employ the uncertainties on scattering fraction \mbox{($f_{\text{scatt}}$ + $\sigma_{\text{scatt}}$)} in our models, to avoid unnecessary confusion we refer to this parameter as $f_{\text{scatt}}$ except where explicitly necessary.

Next, we considered Compton reflection of X-ray photons by the accretion disk. A fraction of the high-energy continuum photons in proximity to the colder material of the accretion disk are thought to be captured and reprocessed, contributing to an X-ray spectral feature around 30 keV known as the Compton reflection ``hump.'' For our models we chose to add the \textsc{pexmon} spectral component, which assumes a simplistic, slab-like geometry of optically thick, cold material. The strength, or scaling factor for reflection ($R$), was then used as an additional parameter in our model. We then multiplied the reflected component by \textsc{zphabs} and \textsc{cabs} to account for attenuation by photoelectric absorption and Compton scattering, respectively.

Finally, we handled attenuation of emission by a more distant, circumnuclear distribution of gas and dust, colloquially referred to as the obscuring torus. For this, we chose to adopt the radiative transfer code of \citet{balokovic2018}, specifically \textsc{borus02}, representing the AGN torus as a uniform density sphere with polar cutouts and a cutoff power-law intrinsic continuum. As the majority of the attenuation in obscured systems is thought to be caused by nuclear material within the gravitational influence of the central SMBH, the high upper limit of the torus column density parameter of \textsc{borus02} allows us to estimate the effects of extremely high levels of obscuration. The \textsc{borus02} model also accounts for Compton reflection by the torus, further contributing to the Compton hump described above.

Starting at $z =$ 0, we increased the line-of-sight column density from 10$^{\text{21}}$ cm$^{-\text{2}}$ to 10$^{\text{25}}$ cm$^{-\text{2}}$ and calculated the change in the observed-to-intrinsic 2--10 keV flux ratio as a function of $N_{\text{H}}$. For this calculation, we adopted the flux at $N_{\text{H}} =$ 10$^{\text{21}}$ cm$^{-\text{2}}$ as the intrinsic, unobscured X-ray flux. The simulated change in the X-ray spectral model flux ratio provides us with a conversion between $N_{\text{H}}$ and $R_{L_{\text{X}}}$, which we adopted for the body of our work. Though \textsc{borus02} obscuration can be set at higher values, the distinction between sources with column densities higher than the CT limit are not well constrained, thus we chose \mbox{$N_{\text{H}}$ = 10$^{\text{25}}$ cm$^{-\text{2}}$} as an upper limit in our models.

We then investigated the effects of different model parameters on $R_{L_{\text{X}}}$, including scattering fraction $f_{\text{scatt}}$, reflection strength $R$, photon index ($\Gamma$), and torus opening angle ($\theta_{\text{OA}}$). Following the prescription above, we constructed spectral models for each set of parameters \{$f_{\text{scatt}}$,\,$R$,\,$\Gamma$,\,$\theta_{\text{OA}}$\}. To account for the extremely low scattering fraction seen in some heavily obscured systems \citep[e.g.,][]{noguchi2010}, we allowed the scattering fraction to vary uniformly between \mbox{$-$2.0 $\leq \sigma_{\text{scatt}} \leq$ 0.5} in steps of 0.5. We chose 0.5$\sigma_{\text{scatt}}$ as an upper limit as sources with little or no obscuration would produce nonphysical results (i.e., total emission scattered greater than 100\%). We allowed the reflection scaling factor to vary between \mbox{0.0 $\leq R \leq$ 2.0} in steps of 0.2, where $R$ = 0.0 corresponds to Compton reflection soley from the back-side of the torus (i.e., no disk reflection). We allowed photon index to vary in steps of 0.1 between \mbox{1.4 $\leq \Gamma \leq$ 2.2} to span the full range observed in AGN samples. Finally, we allowed the opening angle to vary in steps of 10$\degr$, spanning a range from \mbox{10$\degr \leq \theta_{\text{OA}} \leq$ 80$\degr$}.

\begin{figure}
    \plotone{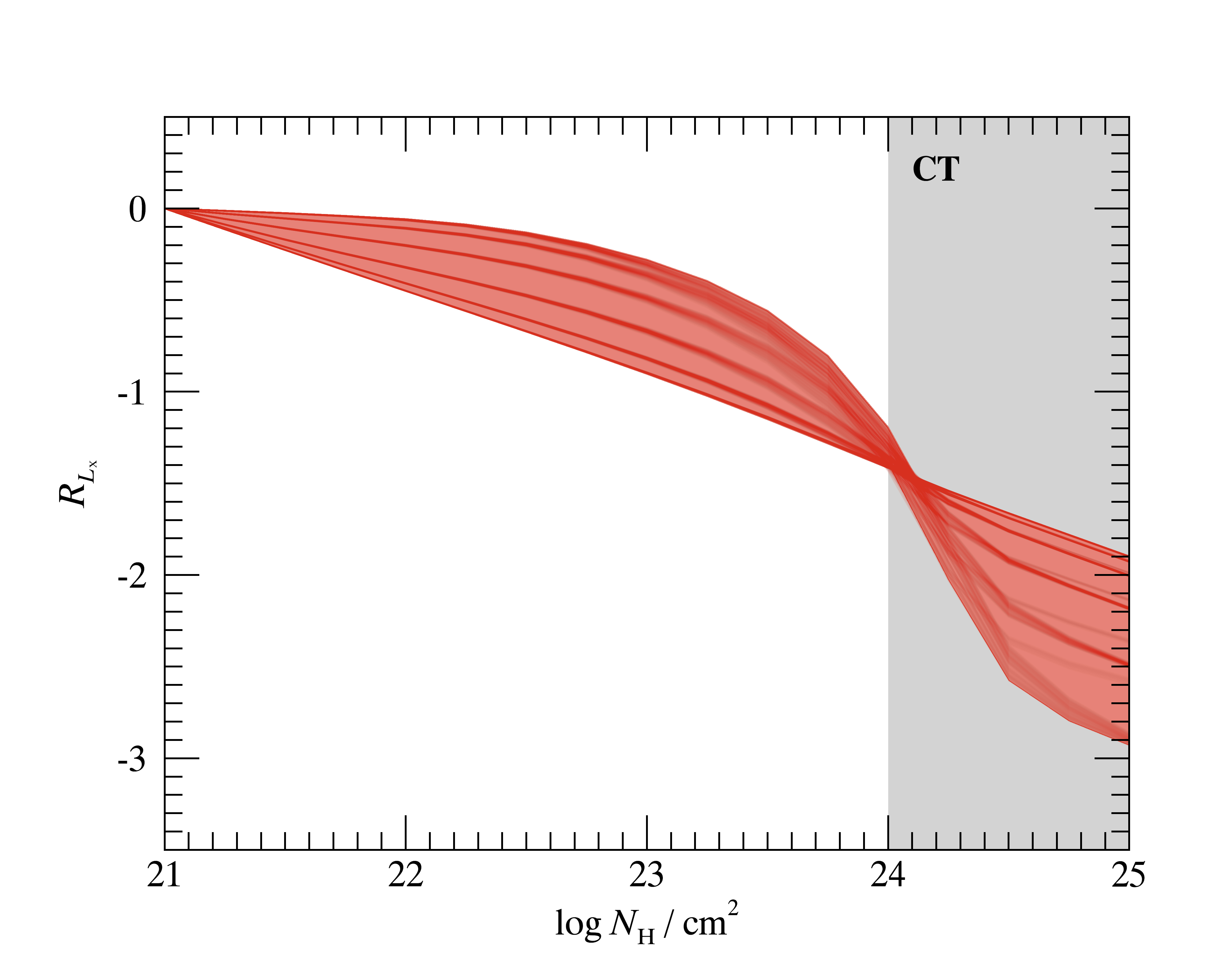}
    \caption{Observed-to-intrinsic X-ray luminosity ratio $R_{L_{\text{X}}}$ as a function of obscuring column density $N_{\text{H}}$. The shaded region of $R_{L_{\text{X}}}$--$N_{\text{H}}$ shown here represents the continuous parameter space inferred from our full suite of X-ray spectral models and parameters \{$f_{\text{scatt}}$,\,$R$,\,$\Gamma$,\,$\theta_{\text{OA}}$\}. Of these parameters we find $f_{\text{scatt}}$ to have the most significant effect on $R_{L_{\text{X}}}$, as seen by the striations, or channels, caused by our choice of step size in $f_{\text{scatt}}$ parameter space. The remaining parameters $R$, $\Gamma$, and $\theta_{\text{OA}}$ are responsible for the narrow scatter within each of the deeper channels attributed to $f_{\text{scatt}}$. The vertical shaded gray region demarcates Compton-thick levels of obscuration \mbox{($N_{\text{H}} >$ 10$^{\text{24}}$ cm$^{-\text{2}}$)}.}
    \label{fig:rxnh}
\end{figure}

\begin{figure}
    \epsscale{1.05}
    \plotone{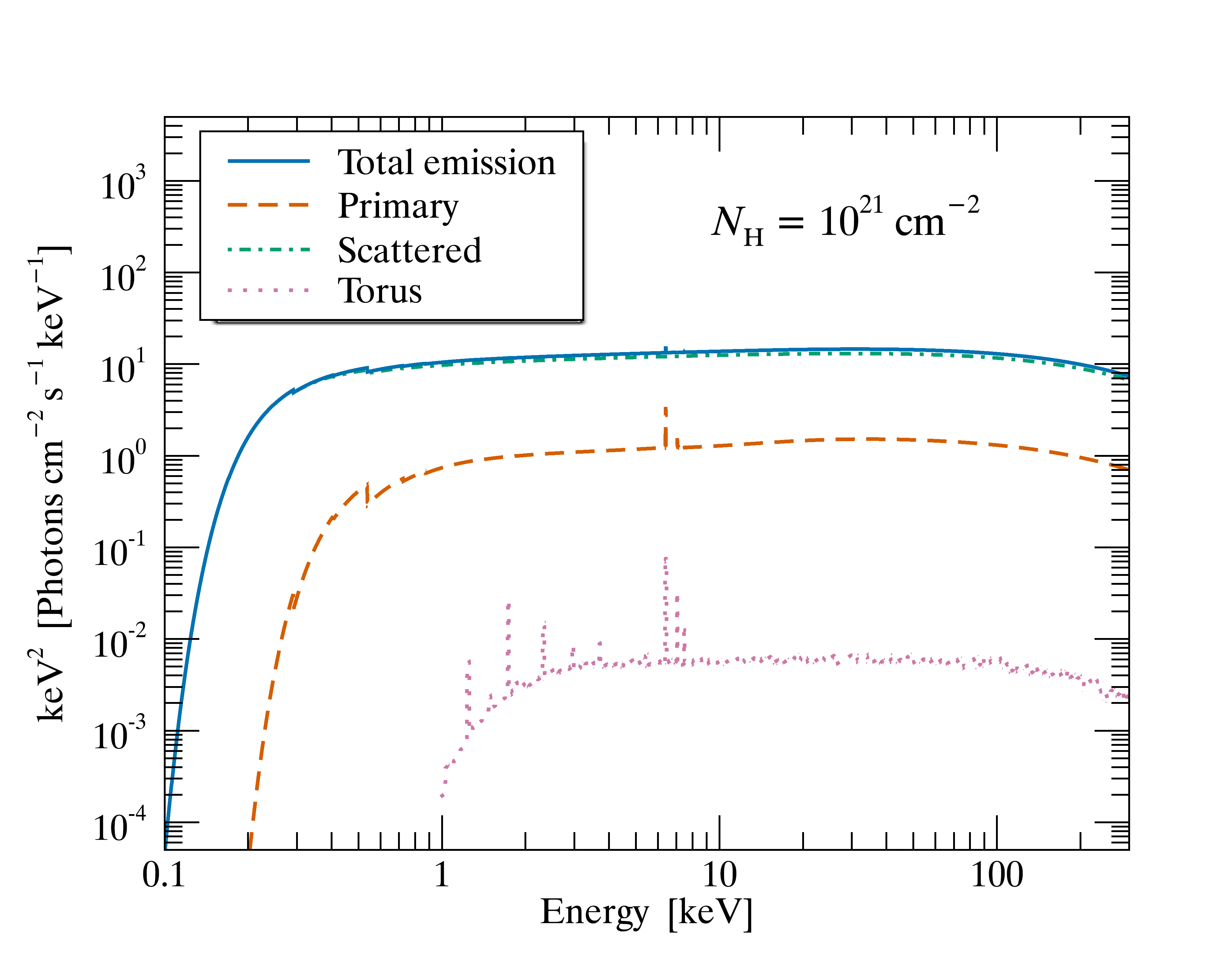}
    \plotone{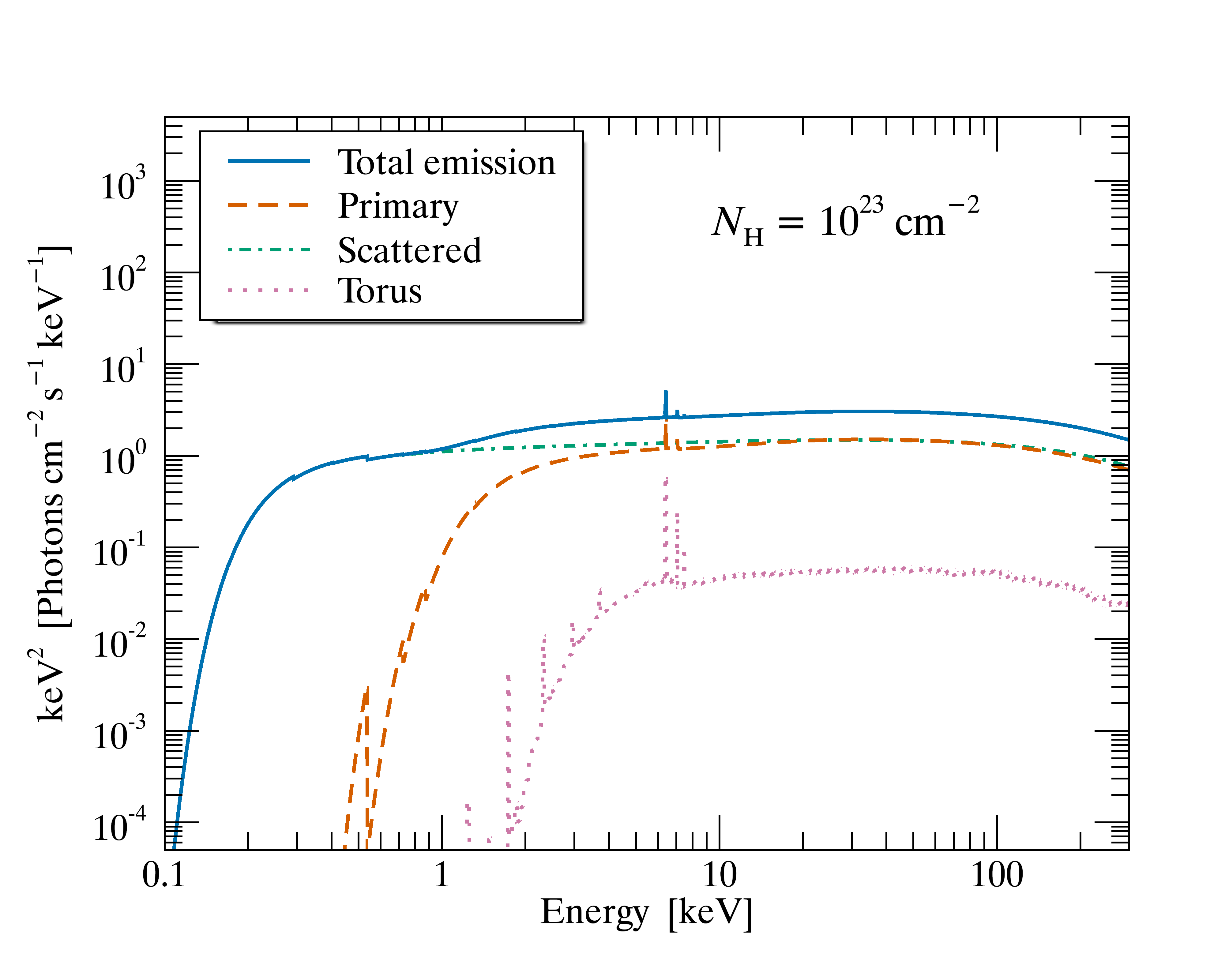}
    \plotone{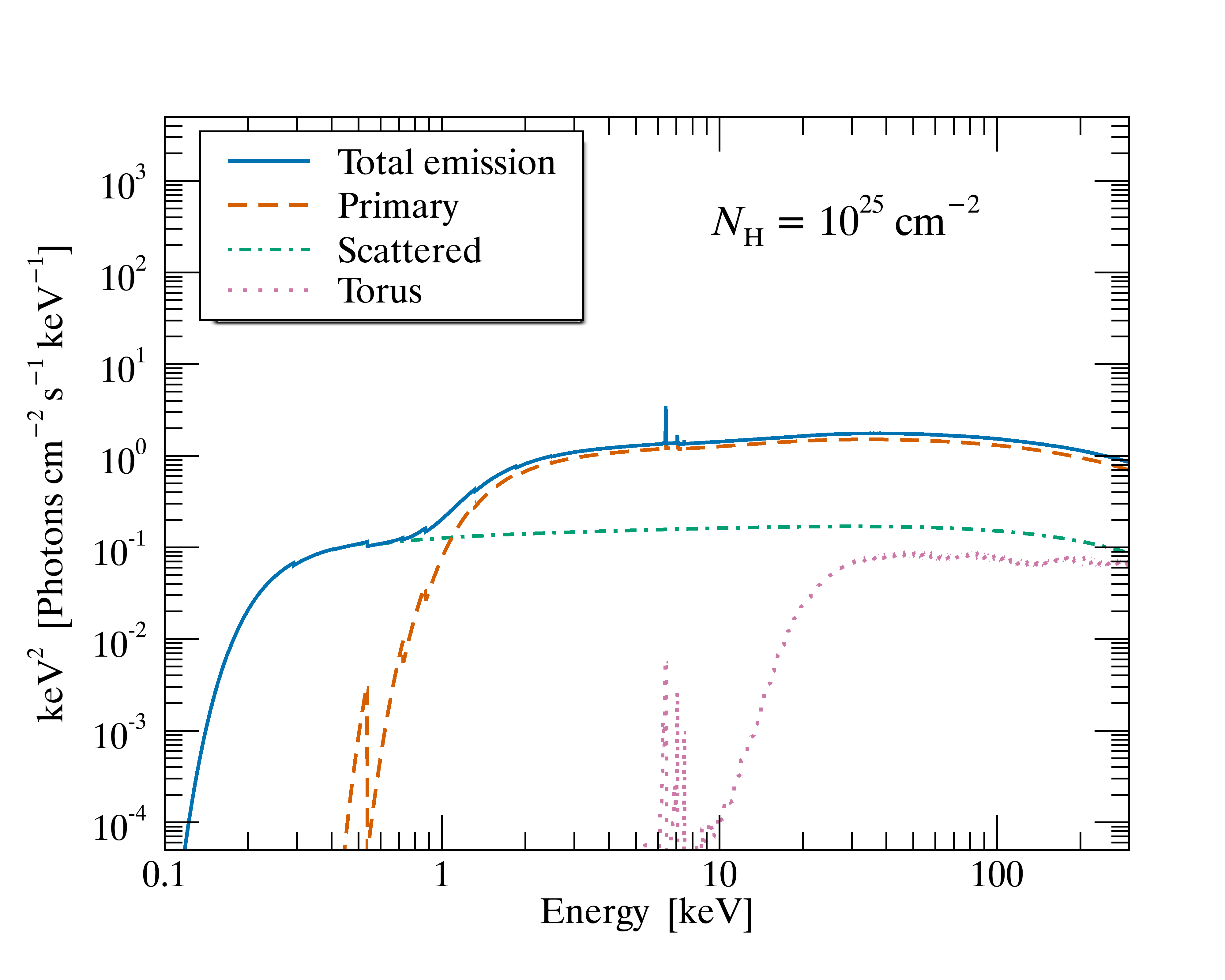}
    \caption{X-ray spectral energy distributions. Variation in the observed-frame X-ray spectra is shown with changes to the absorbing column density $N_{\text{H}}$. To assist with visualization, we show three cases: \mbox{$N_{\text{H}}$ = 10$^{\text{21}}$ cm$^{-\text{21}}$} (top); \mbox{$N_{\text{H}}$ = 10$^{\text{23}}$ cm$^{-\text{2}}$} (middle); and \mbox{$N_{\text{H}}$ = 10$^{\text{25}}$ cm$^{-\text{2}}$} (bottom). The total X-ray SED is shown (solid blue) as well as the individual components: primary power-law continuum disk emission (dashed orange), scattered primary emission (dash-dotted green), and reprocessed torus emission (dotted magenta).}
    \label{fig:xspec}
\end{figure}

The relation between $R_{L_{\text{X}}}$ and $N_{\text{H}}$ from our models is shown in Figure \ref{fig:rxnh}. Of our choice of parameters, variance on $f_{\text{scatt}}$ had the most significant effect on $R_{L_{\text{X}}}$, causing large striations, or channels, to develop in $R_{L_{\text{X}}}$--$N_{\text{H}}$. Aside from $f_{\text{scatt}}$, our other parameters show little to no effect on $R_{L_{\text{X}}}$. Additionally, \citet{ananna2020a} ruled out regions of X-ray spectral parameter space where AGNs could not reliably reproduce the CXB. As these parameters also produce no significant effect on our results, we chose set values moving forward to save computational costs. For the remainder of this work, we adopted $R_{L_{\text{X}}}$ values from our X-ray models where $\Gamma$ = 1.9, $R$ = 0.99, and $\theta_{\text{OA}}$ = 60$\degr$. A subset of our X-ray spectral models are presented in Figure \ref{fig:xspec}.

We then evaluated the observed-frame output of the model for a range of redshifts to cover our sample redshift space \mbox{(0.0 $\leq z \leq$ 0.8)}. At each redshift, we calculated X-ray spectral model fluxes in the observed \mbox{0.5--2} and \mbox{2--7 keV} bands. A comparison between the observed soft and hard fluxes with rest-frame \mbox{2--10 keV} flux provides us with a way to estimate \mbox{0.5--2} and \mbox{2--7 keV} fluxes we would expect to observe for a given model distribution of $N_{\text{H}}$.

%%%%%%%%%%%%%%%%%%%%%%%%%%%%%%%%%%%%%%%%%%%%%%%%%%%%%%%%%%%
%%      Modeling the Compton-thick Fraction
%%%%%%%%%%%%%%%%%%%%%%%%%%%%%%%%%%%%%%%%%%%%%%%%%%%%%%%%%%%
\subsection{Modeling the Compton-thick Fraction} \label{ssec:ctf}
We used the Bayesian ensemble sampler \texttt{emcee} \citep{foreman-mackey2013} to forward model the observed $R_{L_{\text{X}}}$ distribution of our sample and estimate the total fraction of CT AGNs. Using the connection between AGN X-ray flux and nuclear obscuration explored in Section \ref{ssec:xspec}, we created model $N_{\text{H}}$ distributions which we then converted to distributions in $R_{L_{\text{X}}}$. We then used these model distributions $M(R_{L_{\text{X}}})$ to infer the likelihood of observing our sample data.

\begin{figure}
    \plotone{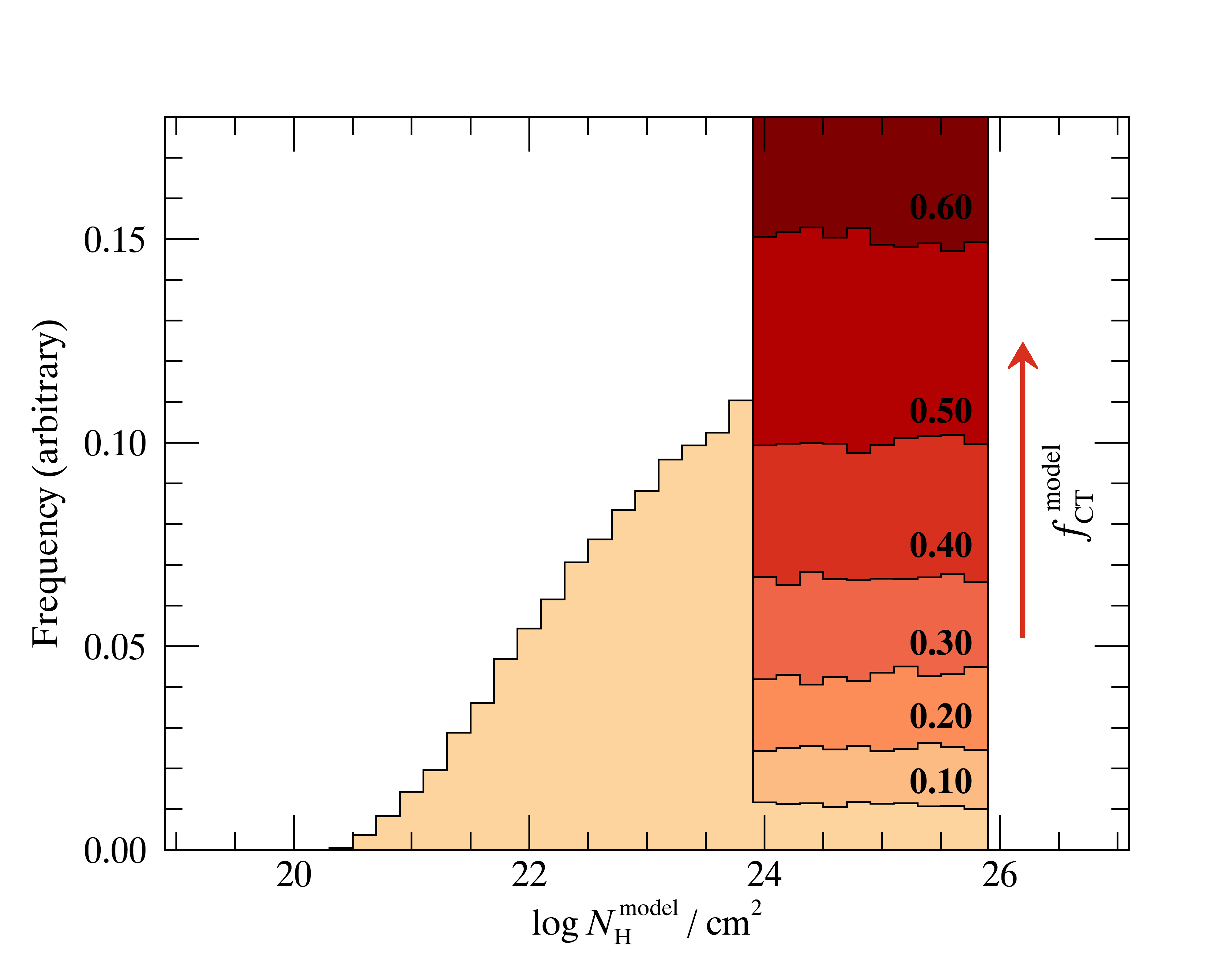}
    \plotone{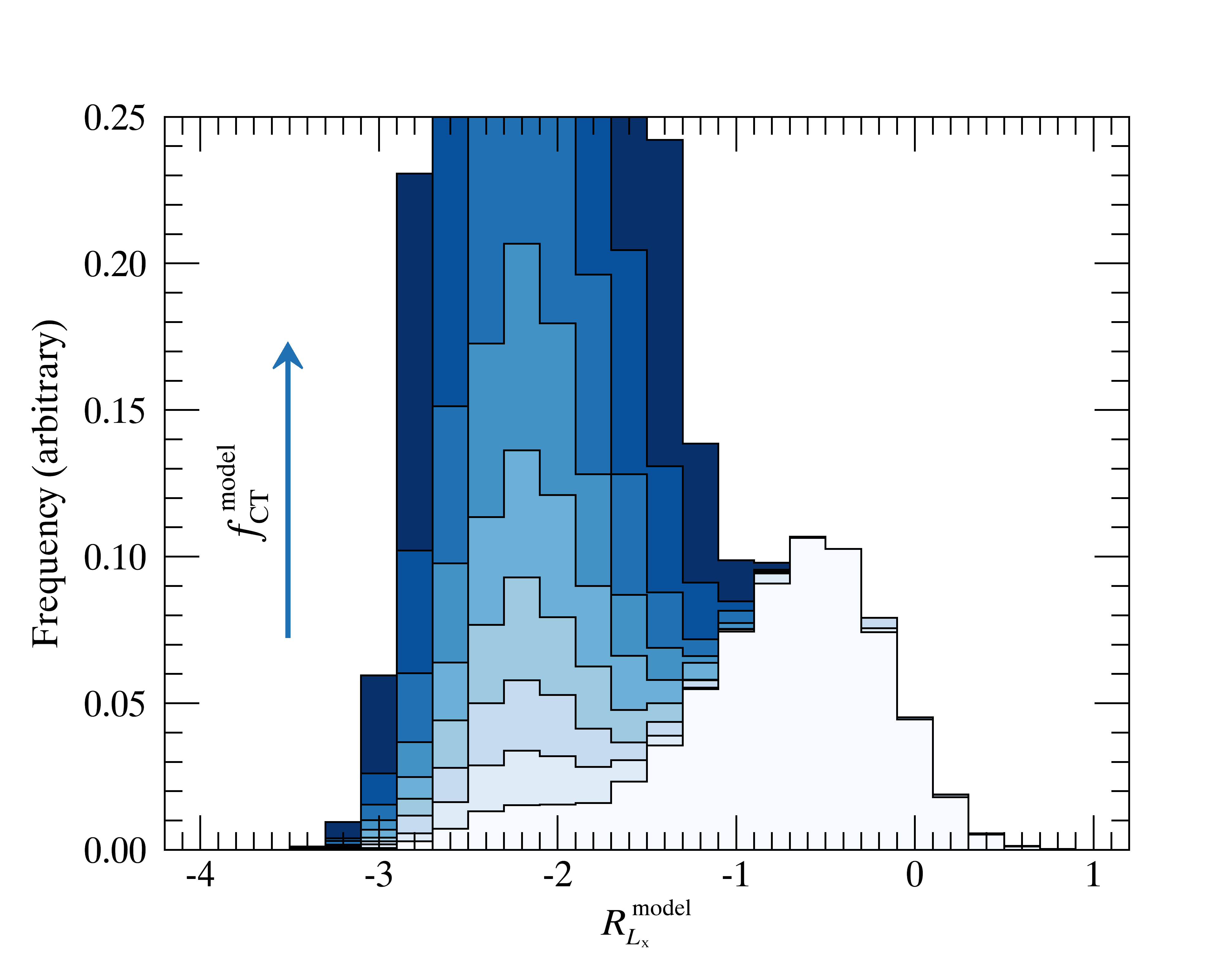}
    \caption{Upper panel: Example $N_{\text{H}}$ model distributions from our simulation, where each model is generated with varying $f_{\text{CT}}$. This figure represents a small subset of possible $N_{\text{H}}$ distributions, presented here in steps of $\Delta f_{\text{CT}}$ = 0.10 with uniformly distributed number of sources at \mbox{$N_{\text{H}} >$ 10$^{\text{24}}$ cm$^{-\text{2}}$} for clarity. Lower panel: Corresponding $R_{L_{\text{X}}}$ distributions.}
    \label{fig:distmod}
\end{figure}

At each step of the MCMC, we used the model parameter $f_{\text{CT}}$ to create a random sample in $N_{\text{H}}$ covering a range of column densities (see Figure \ref{fig:distmod}). For column densities up to 10$^{\text{24}}$ cm$^{-\text{2}}$, we assumed the intrinsic column density distribution of \cite{ricci2017b} for non-blazar AGNs. As C21 focused primarily on the detection of heavily obscured sources, their final sample lacks a significant fraction of true unobscured, \mbox{Type-I AGNs} \mbox{($N_{\text{H}} \approx$ 10$^{\text{20}}$ cm$^{-\text{2}}$)}. Considering the lack of dynamic range in $R_{L_{\text{X}}}$ where \mbox{$N_{\text{H}} <$ 10$^{\text{21}}$ cm$^{-\text{2}}$}, we chose to modify the input $N_{\text{H}}$ distribution by removing all sources with column densities $N_{\text{H}} <$ 10$^{\text{21}}$ cm$^{-\text{2}}$. We further tested our modeling with a different input $N_{\text{H}}$ distributions \citep[i.e., the NuSTAR informed distribution of][]{lansbury2015} and found no significant change in our results, indicating that uncertainties on the fraction of fully unobscured sources does not play a significant role in our modeling and results. We made limited assumptions regarding the number of sources and shape of the underlying distribution for \mbox{$N_{\text{H}} \geq$ 10$^{\text{24}}$ cm$^{-\text{2}}$}. Specifically, we set the upper limit of our model to \mbox{$N_{\text{H}}$ = 10$^{\text{26}}$ cm$^{-\text{2}}$}, as densities approaching this limit have only recently been detected \citep[e.g.,][]{yan2019}. The fraction of heavily obscured sources was allowed to vary between \mbox{0.00 $< f_{\text{CT}} <$ 1.00}, and the Compton scattering fraction $f_{\text{scatt}}$ was allowed to vary within the G21 uncertainties. A flat prior was assumed on both parameters.

Although $f_{\text{CT}}$ and $f_{\text{scatt}}$ are the only free parameters in our simulation, we also considered the shape of the $N_{\text{H}}$ distribution for sources with high obscuration \mbox{($N_{\text{H}} \geq$ 10$^{\text{24}}$ cm$^{-\text{2}}$)}. We allowed the fraction of sources with column densities \mbox{10$^{\text{24}}$--10$^{\text{25}}$ cm$^{-\text{2}}$} and \mbox{10$^{\text{25}}$--10$^{\text{26}}$ cm$^{-\text{2}}$} ($f_{\text{24--25}}$ and $f_{\text{25--26}}$, respectively) to vary, where the combined fractions account for the full number of CT sources in any given model. We performed multiple runs of our simulation to assess the influence a higher fraction of extremely obscured sources \mbox{($f_{\text{24--25}} < f_{\text{25--26}}$)} would have on our parameter estimates. As shown in Figure \ref{fig:param_comp}, we find that varying $f_{\text{24--25}}$ shows a negligible effect on $f_{\text{CT}}$, but shares a positive correlation with $f_{\text{scatt}}$. As the number of heavily obscured sources rises (i.e., $f_{\text{25--26}}$ increases), a slight increase in $f_{\text{scatt}}$ is required to model the data (i.e., the uncertainty on scatter fraction $\sigma_{\text{scatt}}$ moves closer to zero). Variations on  $f_{\text{24--25}}$ and $f_{\text{25--26}}$ produce trivial change in our parameter estimation, therefore we adopt a uniform $N_{\text{H}}$ distribution where $f_{\text{24--25}}$ = $f_{\text{25--26}}$ for our simulation. Additionally, we ran similar tests for a range of parameters \{$R$,\,$\Gamma$,\,$\theta_{\text{OA}}$\}, the results of which are shown in Figures \ref{fig:params_fct} and \ref{fig:params_fscat}.

Our simulation aimed to estimate $f_{\text{CT}}$ from a sample of 540 IR-luminous AGNs, many of which lack X-ray counterparts. To accomplish this, we modeled both the intrinsic number of sources expected in the survey as well as the expected observed sample size, based on both our model and X-ray flux sensitivity limits. We assumed a log-likelihood function of the form

\begin{multline}\label{eq:likelihood}
  \ln \mathcal{L} = 
  \sum^{N_{\text{det}}}_{i=1}\,\ln \int M(R_{L_{\text{X}}})\,\mathcal{N}(R_{L_{\text{X}},i},\,0.23)\,\mathop{dR_{L_{\text{X}}}} \\
  - \int M(R_{L_{\text{X}}})\,\sum^{N_{\text{all}}}_{i=1}\,S_i(R_{L_{\text{X-lim}}})\,\mathop{dR_{L_{\text{X}}}}
\end{multline}
to account for both X-ray detected and non-detected sources. A detailed description of the two terms in our likelihood function follows. A visual representation of $M(R_{L_{\text{X}}})$ and both terms of our likelihood are presented in Figures \ref{fig:mrlx}, \ref{fig:likelihood_first}, and \ref{fig:likelihood_second}, respectively.

The first term of Equation \ref{eq:likelihood} sums over all model distributions for the X-ray detected sources in our sample. Here, $M(R_{L_{\text{X}}})$ represents the normalized model distribution generated at each step in the simulation given a choice of parameters \{$f_{\text{CT}}$,\,$f_{\text{scatt}}$\}. For each X-ray detected source, we create a Gaussian distribution $\mathcal{N}(R_{L_{\text{X}},i},\,\text{0.23})$ centered on the observed $R_{L_{\text{X}}}$ of the source and an uncertainty of 0.23 dex---the intrinsic dispersion in the \mbox{$L_{\text{X}}$--$L_{\text{MIR}}$} relation. The convolution of these two terms generates a probability density of observing a source given the current choice of parameters at each step in the simulation. We then integrated over all possible $R_{L_{\text{X}}}$ and summed over all of the detected sources in the sample.

The second term of Equation \ref{eq:likelihood} acts as a normalizing constant and accounts for all sources in the sample. Again, $M(R_{L_{\text{X}}})$ represents the current model at each step in the simulation, but this time before integrating it is multiplied by the sensitivity function $S_i(R_{L_{\text{X-lim}}})$. To account for observational bias in a flux-limited survey, a complicated calculation of sample incompleteness is generally required. However, we use the fact that $R_{L_{\text{X}}}$ is the logarithmic ratio of X-ray luminosities to our advantage, and instead calculated $R_{L_{\text{X-lim}}}$ using the X-ray flux-limit luminosity of each source (see Equation \ref{eq:rlx}). Using $R_{L_{\text{X-lim}}}$ in this way allows us to estimate whether an object would be detected at any given step in our simulation. As such, the sensitivity function $S_i(R_{L_{\text{X-lim}}})$ is a Heaviside function, shifted by $R_{L_{\text{X-lim}}}$, or
\begin{equation}\label{eq:sensitivity}
  S_i(R_{L_{\text{X-lim}}}) = \Theta(R_{L_{\text{X}}}-R_{L_{\text{X-lim}},i}).
\end{equation}
As before, we convolved $M(R_{L_{\text{X}}})$ with $S_i(R_{L_{\text{X-lim}}})$ and integrated over the full sample. The second term of Equation \ref{eq:likelihood} therefore serves as the expected observed sample size, penalizing our likelihood function where $M(R_{L_{\text{X}}})$ does not produce a comparable number of detected sources.

\begin{figure}
    \epsscale{1.2}
    \plotone{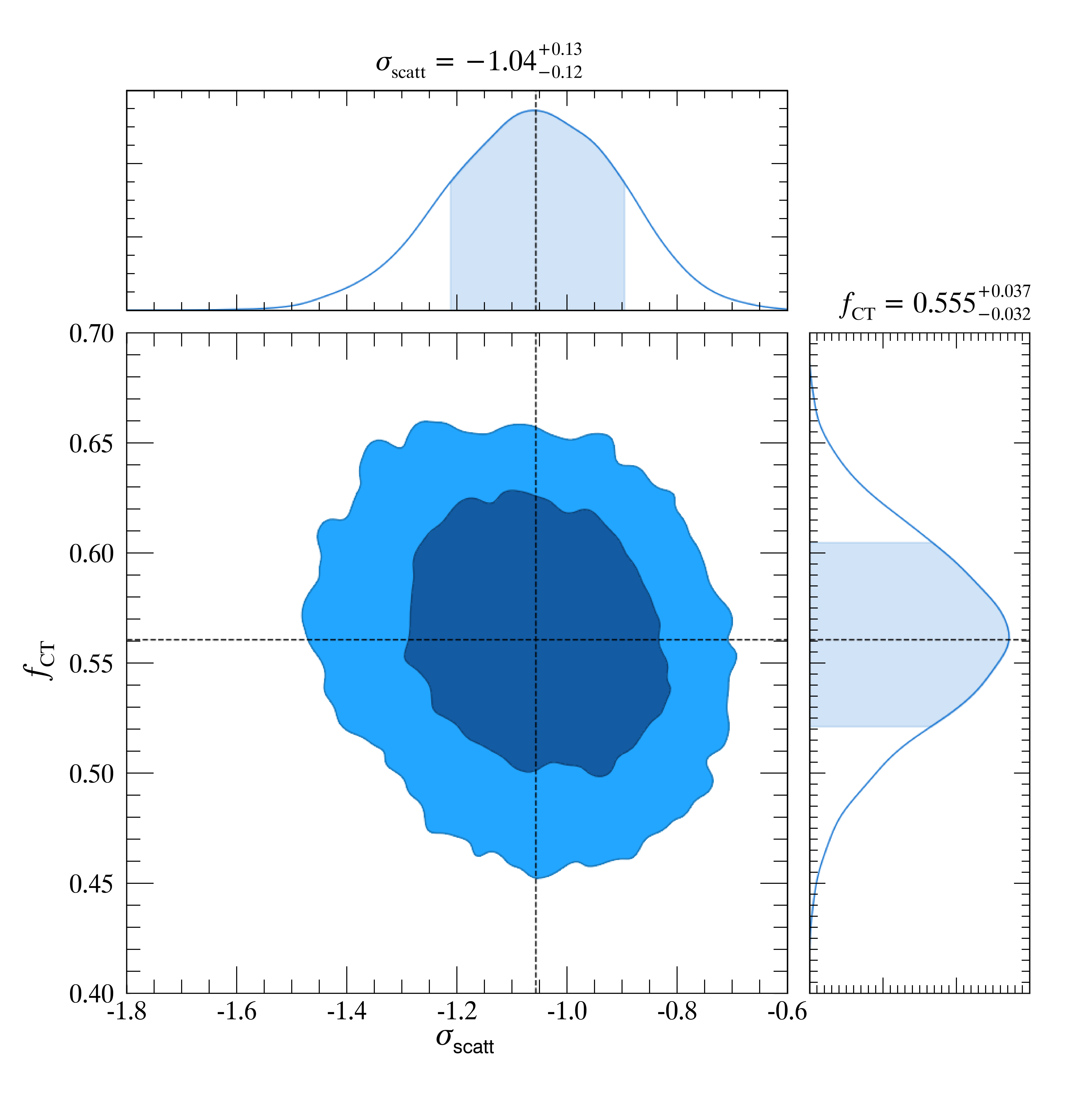}
    \caption{Posterior distributions of our modeling parameters, $f_{\text{CT}}$ and $\sigma_{\text{scatt}}$. The mean and covariance of each parameter is shown alongside its respective distribution.}
    \label{fig:corner}
\end{figure}

The ensemble sampler was run with 80 walkers, each with 1000 steps and a burn-in period of 50 steps. Convergence of the walkers was assessed within \texttt{emcee} using the Gelman-Rubin statistic. The posterior distributions of our parameters were estimated by combining the results of all walkers using the ensemble average (see Figure \ref{fig:corner}).

%%%%%%%%%%%%%%%%%%%%%%%%%%%%%%%%%%%%%%%%%%%%%%%%%%%%%%%%%%%
%%
%%      Results and Conclusions
%%
%%%%%%%%%%%%%%%%%%%%%%%%%%%%%%%%%%%%%%%%%%%%%%%%%%%%%%%%%%%
\section{Results and Conclusions} \label{sec:analysis}
The posterior mean $f_{\text{CT}}$ is estimated to be $\text{0.555}^{+\text{0.037}}_{-\text{0.032}}$, which suggests that at least half of all MIR-selected AGNs out to \mbox{$z \leq$ 0.8} are heavily obscured, with hydrogen column densities on the order of \mbox{{$\sim$}10$^{\text{24}}$ cm$^{-{\text{2}}}$} or higher. Uncertainties on $f_{\text{CT}}$ were inferred from our modeling procedure, which accounted for a variety of parameters (e.g., scattering fraction) and assumptions (e.g., choice of input $N_{\text{H}}$ distribution). Our results are in remarkable agreement with A19 that estimates a luminosity dependent CT fraction on the order of 0.50--0.56 at comparable redshifts ($z \leq$ 1 compared to $z \leq$ 0.8 in this work). Most notable is our arrival at nearly identical results using completely independent methods---we make no use of AGN XLF, nor adjust our methodology to account for the integrated CXB.

\begin{figure*}
    \plotone{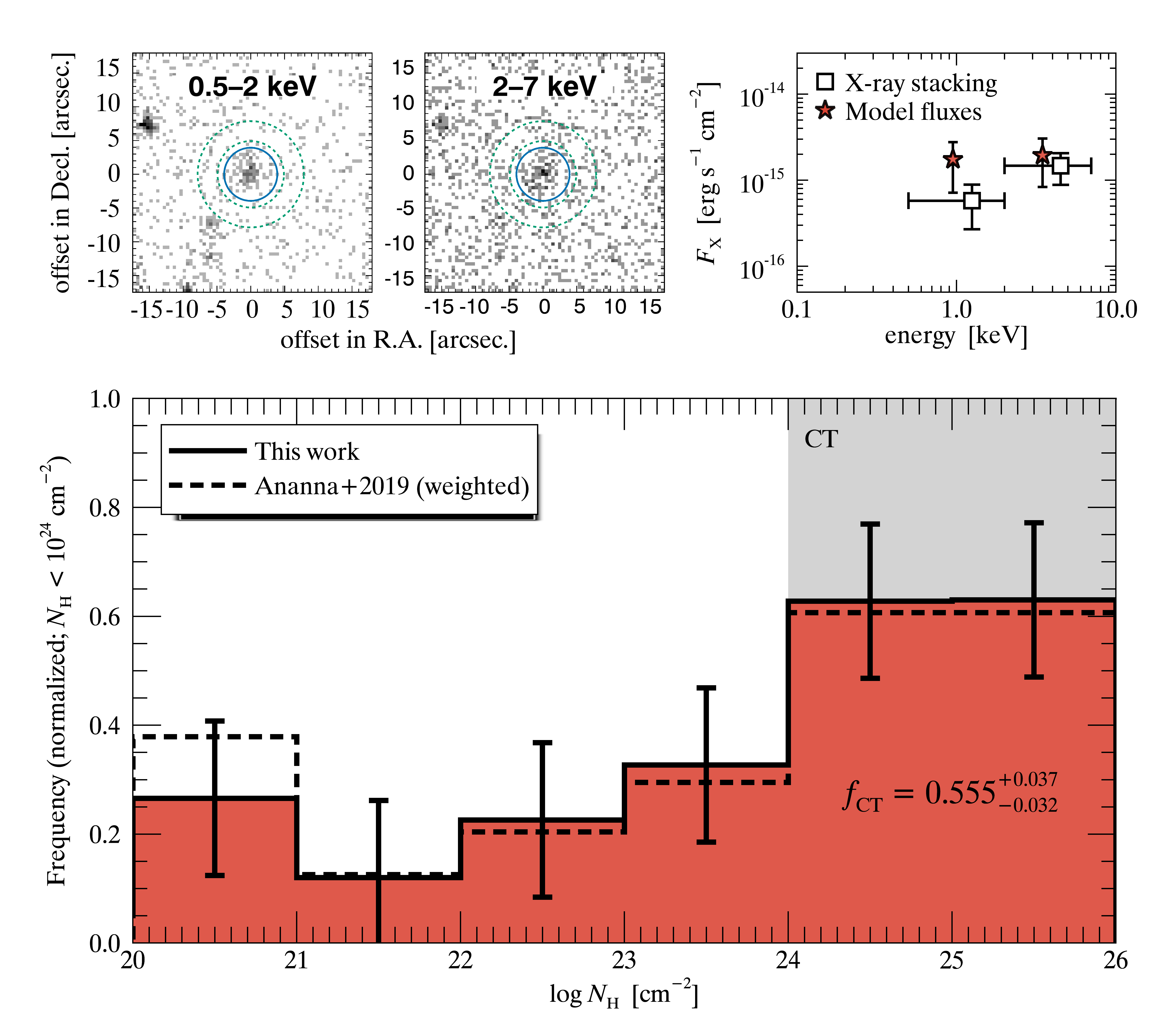}
    \caption{Modeling results. Top left: Soft (0.5--2 keV) and hard (2--7 keV) band stacked images for the X-ray non-detected sources in our sample using \textsc{StackFast} \citep{ananna2023}. Source extraction regions are represented by the solid blue circles, while background subtraction regions are depicted by the dashed green annuli. Top right: Comparison of the X-ray stacked fluxes (squares) to estimates on observed fluxes from our modeling (stars). Model fluxes were plotted with horizontal offsets for visual clarity. Bottom panel: $N{\text{H}}$ distribution with predicted CT fraction of 0.555$^{+\text{0.037}}_{-\text{0.032}}$ from our simulation (solid line). The estimated $N_{\text{H}}$ distribution of A19 is shown for reference (dashed line), where the synthesis model estimates are represented as a weighted average of our sample sources with low ($\log L_{\text{X}}(L_{\text{MIR}})/\text{erg s}^{-1} <$ 43.6) and high ($\log L_{\text{X}}(L_{\text{MIR}})/\text{erg s}^{-\text{1}} \geq$ 43.6) intrinsic X-ray luminosity. The vertical shaded gray region demarcates Compton-thick levels of obscuration \mbox{($N_{\text{H}} >$ 10$^{\text{24}}$ cm$^{-\text{2}}$)}. Sources with $N_{\text{H}} <$ 10$^{\text{20}}$ cm$^{-{\text{2}}}$ (due to scatter converting between $R_{L_{\text{X}}}$ and $N_{\text{H}}$) were reassigned to $N_{\text{H}}$ = 10$^{\text{20}}$ cm$^{-{\text{2}}}$ for clarity.}
    \label{fig:nhdist}
\end{figure*}

We further used our results to infer the distribution of AGN obscuration, including contribution from unidentified CT AGNs. To derive a comprehensive estimate, we used the intrinsic $N_{\text{H}}$ distribution of \citet{ricci2017b}---complete with unobscured sources (i.e., \mbox{$N_{\text{H}} <$ 10$^{\text{21}}$ cm$^{-\text{2}}$})---as our input distribution. We generated 10,000 $N_{\text{H}}$ distributions following the prescription detailed in Section \ref{sec:modeling}, but instead fix the CT fraction to \mbox{$f_{\text{CT}}$ = $\text{0.555}^{+\text{0.037}}_{-\text{0.032}}$}. We then converted $N_{\text{H}}$ to $M(R_{L_{\text{X}}})$ assuming a Gaussian distribution $\mathcal{N}(-\text{1},\,\text{0.8})$ for the uncertainties on $f_{\text{scatt}}$ and the same choice of spectral parameters described in Section \ref{ssec:xspec}. Each simulated source was then randomly assigned to an object in our sample. We compared $M(R_{L_{\text{X}}})$ to $R_{L_{\text{X-lim}}}$ and flagged each simulated source as ``detected'' where \mbox{$M(R_{L_{\text{X}}}) \geq R_{L_{\text{X-lim}}}$} and ``non-detected'' where \mbox{$M(R_{L_{\text{X}}}) < R_{L_{\text{X-lim}}}$}. Finally, we averaged and normalized \mbox{($N_{\text{H}} <$ 10$^{\text{24}}$ cm$^{-\text{2}}$)} over all $N_{\text{H}}$ distributions. Our results modeling the $N_{\text{H}}$ distribution of MIR-selected AGNs is presented in the lower panel of Figure \ref{fig:nhdist}.

We then compared our inferred $N_{\text{H}}$ distribution to that of A19. As their results are luminosity dependent, we averaged over both A19 $N_{\text{H}}$ distributions, weighted by the fraction of our sample sources in each respective luminosity bin (0.21 and 0.79; split on \mbox{log $L_{\text{X}}$/erg s$^{-{\text{1}}}$ = 43.6}). Comparing these results in Figure \ref{fig:nhdist}, we find a striking similarity between our work and that of A19. Though previous estimates of the CT fraction using the AGN XLF have sometimes yielded drastically different results, our determination of the CT fraction and estimated $N_{\text{H}}$ distribution is in close agreement with A19, using completely different measurement techniques and methodologies.

\begin{deluxetable*}{lrrrrrrr}[h]
\tablecolumns{8}
\tablecaption{X-ray stacking results and model comparison.\label{tab:stacking}}
\tablehead{
\colhead{} & \colhead{$t_{\text{exp}}$} & \colhead{Energy} & \colhead{$N_{\text{src}}$} & \colhead{$N_{\text{bkg}}$} & \colhead{$N_{\text{net}}$} & \colhead{$F_{\text{X}}^{\text{stack}}$} & \colhead{$F_{\text{X}}^{\text{model}}$} \\
\colhead{} & \colhead{[Ms]} & \colhead{[keV]} & \colhead{} & \colhead{} & \colhead{} & \colhead{[erg\,s$^{-1}$\,cm$^{-2}$]} & \colhead{[erg\,s$^{-1}$\,cm$^{-2}$]}}
\startdata
WISE AGN (X-ray non-det.) & 1.49 & 0.5--2 &  740 &  634 & 106 & 5.78$\pm$3.09$\times10^{-16}$ & 1.74$\pm$1.02$\times10^{-15}$ \\
                          &      &   2--7 & 1328 & 1219 & 109 & 1.47$\pm$0.59$\times10^{-15}$ & 1.94$\pm$1.11$\times10^{-15}$ \\
\enddata
\tablecomments{X-ray stacking results for energy ranges of \mbox{0.5--2} and \mbox{2--7 keV} with \textsc{StackFast}. For each energy range, we present exposure time, photon counts (source, background, and net), and stacked flux estimates. The final column depicts mean estimates of observed fluxes from our modeling.}
%\tablerefs{}
\end{deluxetable*}

Finally, we considered whether $f_{\text{CT}}$ and the $N_{\text{H}}$ distribution predicted by our modeling would produce similar soft and hard X-ray fluxes comparable to measurements obtained from X-ray stacking. We performed an X-ray stacking analysis of sources in Chandra fields using \textsc{stackfast} \citep{ananna2023}. To ensure accuracy, we restricted our comparison to sources in fields observed by Chandra, rather than XMM or NuSTAR, as the high angular resolution of Chandra allows for more robust X-ray stacking measurements. Of the 540 sources in our dataset, 151 sources lie within observed Chandra fields (60 X-ray detected; 91 X-ray non-detected). For each of our X-ray non-detected sources we extracted photon counts and exposure times in both soft (0.5--2 keV) and hard (2--7 keV) energies. We computed fluxes using count-rate-to-flux ratios characteristic of our X-ray spectral models (see Section \ref{ssec:xspec}), and assumed a power law spectrum with photon index $\Gamma$ = 1.8 and typical Galactic absorption ($N_{\text{H}}$ = 10$^{\text{20}}$ cm$^{-\text{2}}$) as is typical of observed AGN X-ray spectra. We adopted count-rate-to-flux conversion factor of 3.61$\times$10$^{-\text{12}}$ erg cm$^{-\text{2}}$ count$^{-\text{1}}$ and 7.44$\times$10$^{-\text{12}}$ erg cm$^{-\text{2}}$ count$^{-\text{1}}$ for 0.5--2 and 2--7 keV, respectively. Fluxes calculated from our count-rate-to-flux conversion are representative of Chandra ACIS-I Cycle 12 responses---the current cycle for the majority of the observations in our stacking analysis. Uncertainties on the stacked fluxes were estimated by bootstrap resampling of input sources. The calculated uncertainties therefore reflect the distribution of input fluxes and are generally larger than Poisson uncertainties. Although our choice of X-ray spectral models in Section \ref{ssec:xspec} includes additional components, the spectral shape chosen for our count-rate-to-flux conversion is consistent in the energy range considered in this work (2--10 keV) and was deemed sufficient for our analysis.

To estimate the observed X-ray fluxes from our model, we randomly assigned each simulated sources previously flagged as non-detected to one of the 91 X-ray non-detected Chandra sources in our sample. We extracted rest-frame 2--10 keV luminosities for each model source using the combination of $R^{\text{model}}_{L_{\text{X}}}$ and $L_{\text{X}}(L_{\text{MIR}})$. We then converted the rest-frame 2--10 keV luminosities to observed 0.5--2 and 2--7 keV fluxes using the X-ray spectral conversions discussed in Section \ref{ssec:xspec}. To account for possible contamination in the data, we randomly set the X-ray flux for 10\% of all model sources to zero. We chose to remove 10\% of model sources to match the WISE AGN R90 catalog \citep{assef2018}, which assesses 90\% reliability in their AGN selection methods. Though recent work has suggested that the level of contamination may be higher \citep[e.g.,][]{hainline2016b,lamassa2019}, we found that varying the fraction of contamination between 5\%--20\% had a negligible effect on our results. We repeated this process 100 times and calculated the mean flux over all model sources.

The results of both our X-ray stacking and model estimates are presented in Table \ref{tab:stacking}, while a side-by-side visual comparison is also shown in the top panel of Figure \ref{fig:nhdist}. We find our model fluxes to be in good agreement with the X-ray stacking results to within 0.48 dex at soft energies (0.5--2 keV) and 0.12 dex at hard energies (2--7 keV). While our hard X-ray fluxes are in excellent agreement with our stacking results, our soft X-ray fluxes estimates are three times higher than our stacking results. An increase in X-ray photons at soft energies could possibly be due to overestimation of the reflection strength parameter $R$ used in our spectral models. In fact, one postulated explanation of the soft X-ray excess found in nearly half of all nearby AGNs is upscattered X-ray photons reflecting back off of the accretion disk. Though this model may be losing favor \citep[e.g.,][]{done2007,ursini2020}, it stands that reflection does contribute to an increase in soft X-rays. The discrepancy between the soft X-ray fluxes of our model and stacking may be caused by a relatively high amount of Compton scattering, generally attributed to Type-I AGNs.

We also considered whether the X-ray selection of targeted sources may bias our results. To test this we removed targets close to field centers (at both 0.5$'$ and 1$'$) and reran our analysis to track any changes. In the end, we found no appreciable differences in our results. We find that the fraction of detected sources actually increases linearly with distance from field center. Because of this, we do not believe there is any significant bias introduced by combining X-ray targeted sources with serendipitous sources. It is possible that detection of point sources in the vicinity of targeted bright X-ray objects could be limited by diffuse emission, but this is not something we can currently account for.

%%%%%%%%%%%%%%%%%%%%%%%%%%%%%%%%%%%%%%%%%%%%%%%%%%%%%%%%%%%
%%
%%      Discussion
%%
%%%%%%%%%%%%%%%%%%%%%%%%%%%%%%%%%%%%%%%%%%%%%%%%%%%%%%%%%%%
\section{Discussion} \label{sec:discussion}
Our approach differs from previous work in distinct ways: modeling of the observed $R_{L_{\text{X}}}$ distribution (i.e., a measurable proxy for nuclear obscuration) in lieu of assumptions on $N_{\text{H}}$ distribution; the choice of generalized X-ray spectral models, accounting for fewer parameters, degrees of freedom, and potential degeneracies; constraints from X-ray stacking of individual sources rather than an integrated spectrum, such as the CXB. Whereas synthesis models may fit obscured and unobscured AGNs separately to estimate the overall XLF, our technique relies solely on direct observables (i.e., X-ray and infrared observations) to infer the overall distribution of CT AGNs from simulated $N_{\text{H}}$ distributions. Our approach of forward modeling the observed $R_{L_{\text{X}}}$ distribution allows us to deduce the fraction of CT AGNs without assumptions and uncertain completeness corrections on the number of sources or the shape of the $N_{\text{H}}$ distribution for heavily obscured sources.

There are a number of broad implications for a high fraction of CT AGNs, but one direct consequence involves BH accretion physics. A significant increase in the number of heavily obscured AGNs, missing in typical X-ray observations, may signify an increase in the average cosmic accretion efficiency driving BH growth \citep[i.e.,][]{comastri2015}. The BH mass function, which provides a record of the mass of SMBHs in the local universe, in turn provides information on the accretion history of SMBHs. An increase in the integrated energy density due to AGNs would necessitate an increase in the efficiency of accretion, under the assumption that the empirically derived BH mass function is a statistically complete relation \citep[i.e.,][]{blecha2016}. As the AGN community converges on the true underlying population of actively accreting SMBHs in the Universe, we can take further steps to painting a more complete picture of galaxy and AGN co-evolution and the role AGNs play in other observables, such as the CXB.

Furthermore, significant efforts have been made to understand the importance of major mergers on the activity and duty cycle of AGNs. Though the significance of mergers as a primary driver of SMBH growth remains a matter of active investigation, clear evidence is seen for an increase in the fraction of disturbed host morphologies in heavily obscured and CT AGNs \citep[e.g.,][]{kocevski2015}. While mergers are predicted to provide gas which may obscure AGNs at host galaxy scales \citep[e.g.,][]{blecha2018}, the results presented in this work deal primarily with nuclear attenuation. A follow-up study on the host morphologies of our sample, which include many CT AGNs yet undetected at X-ray energies, could provide valuable insight on potential biases due to AGN selection methods in such studies. If the ``missing'' CT AGNs in our sample are found to have particular galaxy morphologies, this would provide further evidence for a scenario in which CT AGN can be obscured by material on scales beyond the central torus and may represent a distinct phase of SMBH growth \citep{sanders1988}.

Future work may attempt to extrapolate our estimate of the Compton-thick fraction from a MIR-selected sample to the full AGN population. A full accounting of observational biases must be considered to ensure a complete working sample of AGNs collected from different selection techniques at various wavelengths.

%%%%%%%%%%%%%%%%%%%%%%%%%%%%%%%%%%%%%%%%%%%%%%%%%%%%%%%%%%%
%%
%%      Acknowledgments
%%
%%%%%%%%%%%%%%%%%%%%%%%%%%%%%%%%%%%%%%%%%%%%%%%%%%%%%%%%%%%
\begin{acknowledgments}
We thank the referee for exceptionally thoughtful and constructive comments which led to a substantial improvement of this work.

This research has made use of X-ray data from the following: the NuSTAR mission, a project led by the California Institute of Technology, managed by the Jet Propulsion Laboratory, and funded by the National Aeronautics and Space Administration (NASA); the Chandra Source Catalog 2.0, provided by the Chandra X-ray Center (CXC) as part of the Chandra Data Archive; XMM-Newton, a European Space Agency (ESA) science mission with instruments and contributions directly funded by ESA Member States and NASA. This research also made use of data products from the Wide-field Infrared Survey Explorer, which is a joint project of the University of California, Los Angeles, and the Jet Propulsion Laboratory/California Institute of Technology, and funded by NASA. This work is also based in part on NIR data obtained as part of the UKIRT Infrared Deep Sky Survey, as well as data products from the Two Micron All Sky Survey, which is a joint project of the University of Massachusetts and the Infrared Processing and Analysis Center/California Institute of Technology, funded by NASA and the National Science Foundation.

This research also made use of software provided by the High Energy Astrophysics Science Archive Research Center (HEASARC), which is a service of the Astrophysics Science Division at NASA/GSFC and the High Energy Astrophysics Division of the Smithsonian Astrophysical Observatory.

C.M.C. acknowledges support from the John Templeton Foundation through LSSTC Catalyst Fellowship grant number 62192. C.J.C., R.C.H., and T.T.A. acknowledge support from the NSF through CAREER award number 1554584. R.C.H. and T.T.A. acknowledge support from NASA through ADAP grant number 80NSSC19K0580. T.T.A. also acknowledges support from NASA ADAP grant number 80NSSC23K0557. R.J.A. was supported by FONDECYT grant number 1191124 and 1231718, and by ANID BASAL project FB210003.
\end{acknowledgments}

\facilities{Chandra, GALEX, MMT, NuSTAR, SDSS, UKIRT, WISE, XMM, 2MASS}

\software{\texttt{emcee}: The MCMC Hammer \citep{foreman-mackey2013}, \textsc{xspec} \citep{arnaud1996}}

%%%%%%%%%%%%%%%%%%%%%%%%%%%%%%%%%%%%%%%%%%%%%%%%%%%%%%%%%%%
%%
%%      Appendix
%%
%%%%%%%%%%%%%%%%%%%%%%%%%%%%%%%%%%%%%%%%%%%%%%%%%%%%%%%%%%%
\clearpage
\appendix
%%%%%%%%%%%%%%%%%%%%%%%%%%%%%%%%%%%%%%%%%%%%%%%%%%%%%%%%%%%
%%      Appendix: Additional Parameter Tests
%%%%%%%%%%%%%%%%%%%%%%%%%%%%%%%%%%%%%%%%%%%%%%%%%%%%%%%%%%%
\restartappendixnumbering
\section{Additional Parameter Tests}
\begin{figure}[h]
    \epsscale{0.8}
    \plottwo{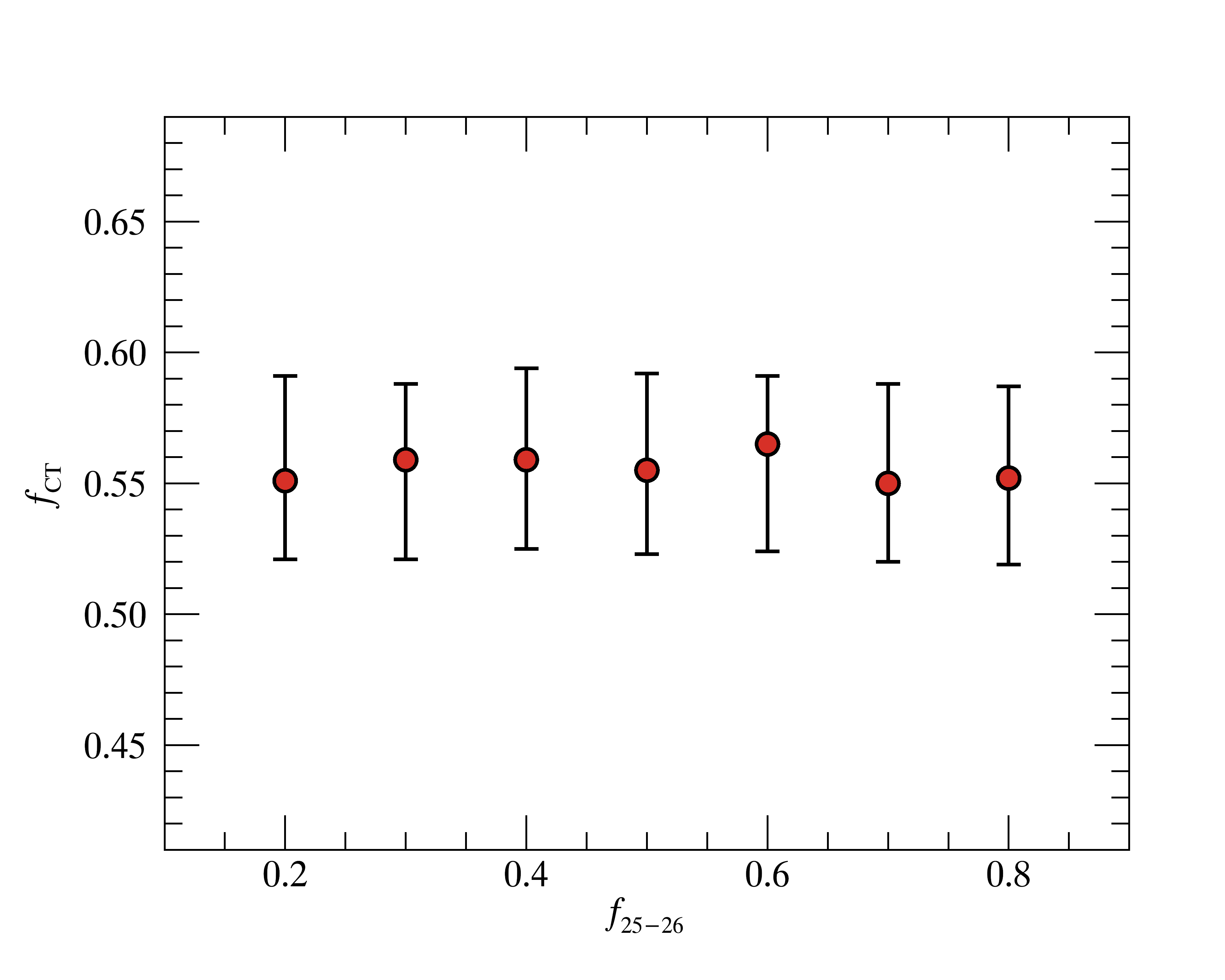}{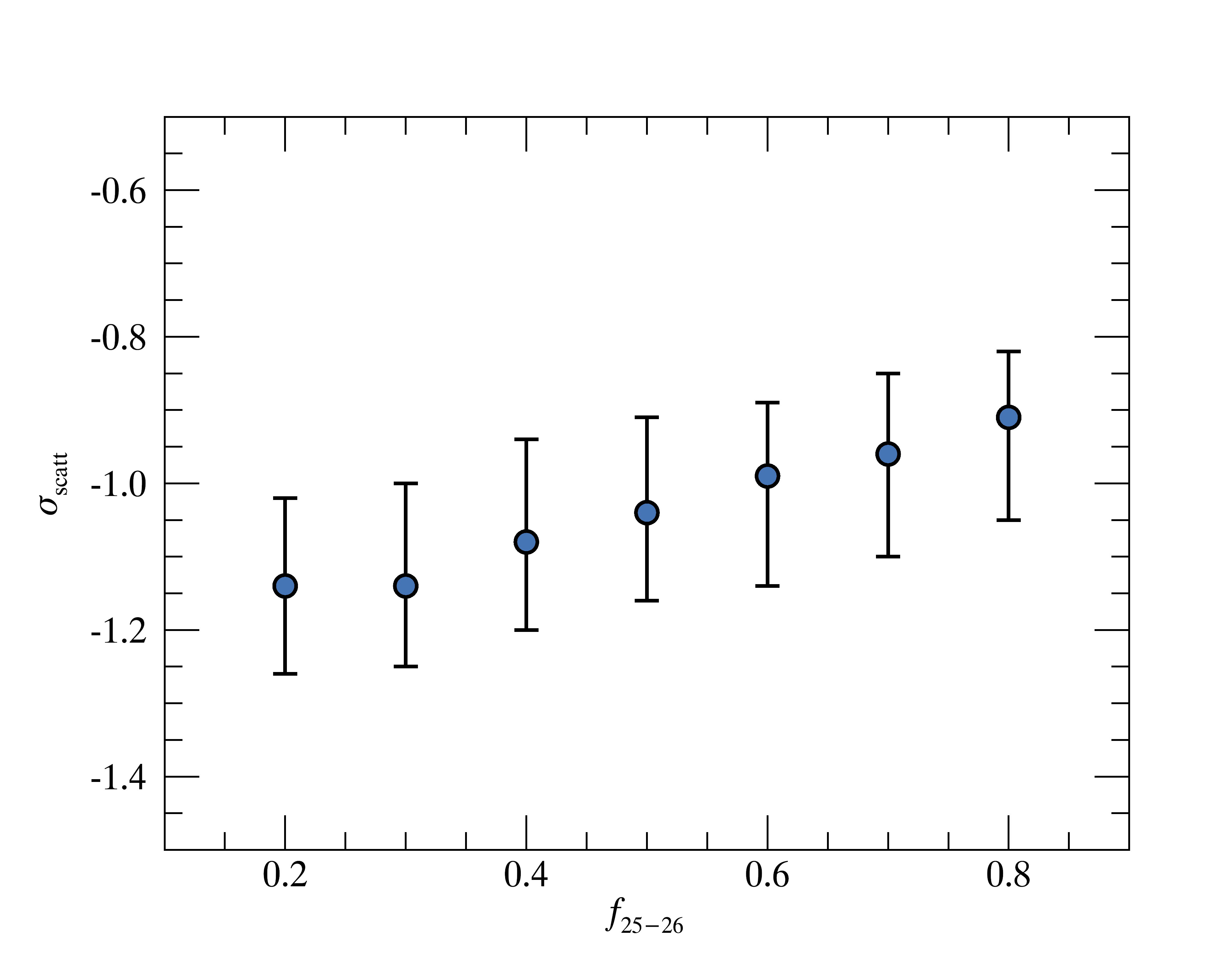}
    \caption{Effects on parameter estimates for a variable distribution of CT sources. Left: Estimates on $f_{\text{CT}}$ for increasing fraction of heavily obscured sources $f_{\text{25--26}}$. Right: Similar estimates on $\sigma_{\text{scatt}}$.}
    \label{fig:param_comp}
\end{figure}
\vspace{-7mm}
\begin{figure}[h]
    \epsscale{0.8}
    \plottwo{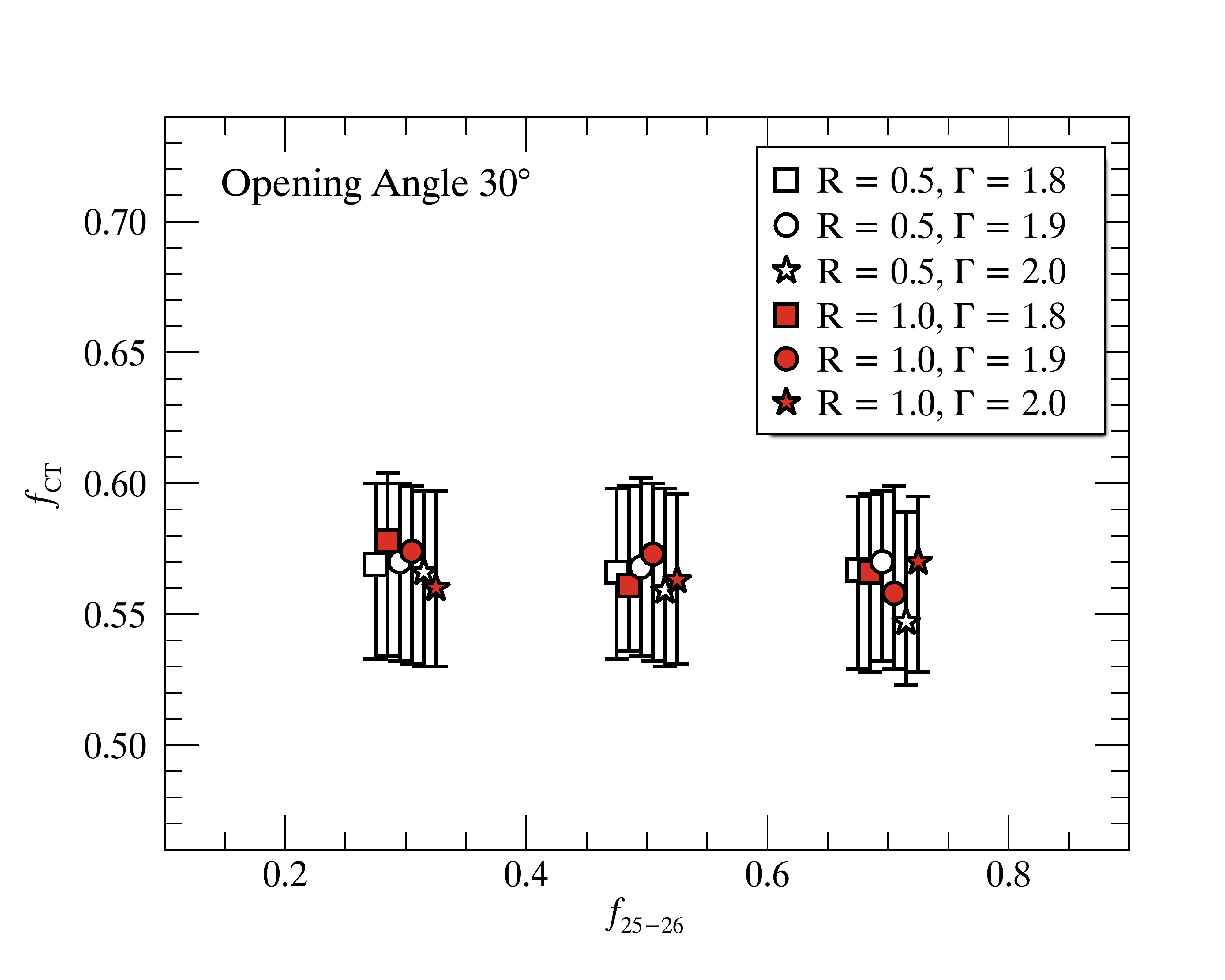}{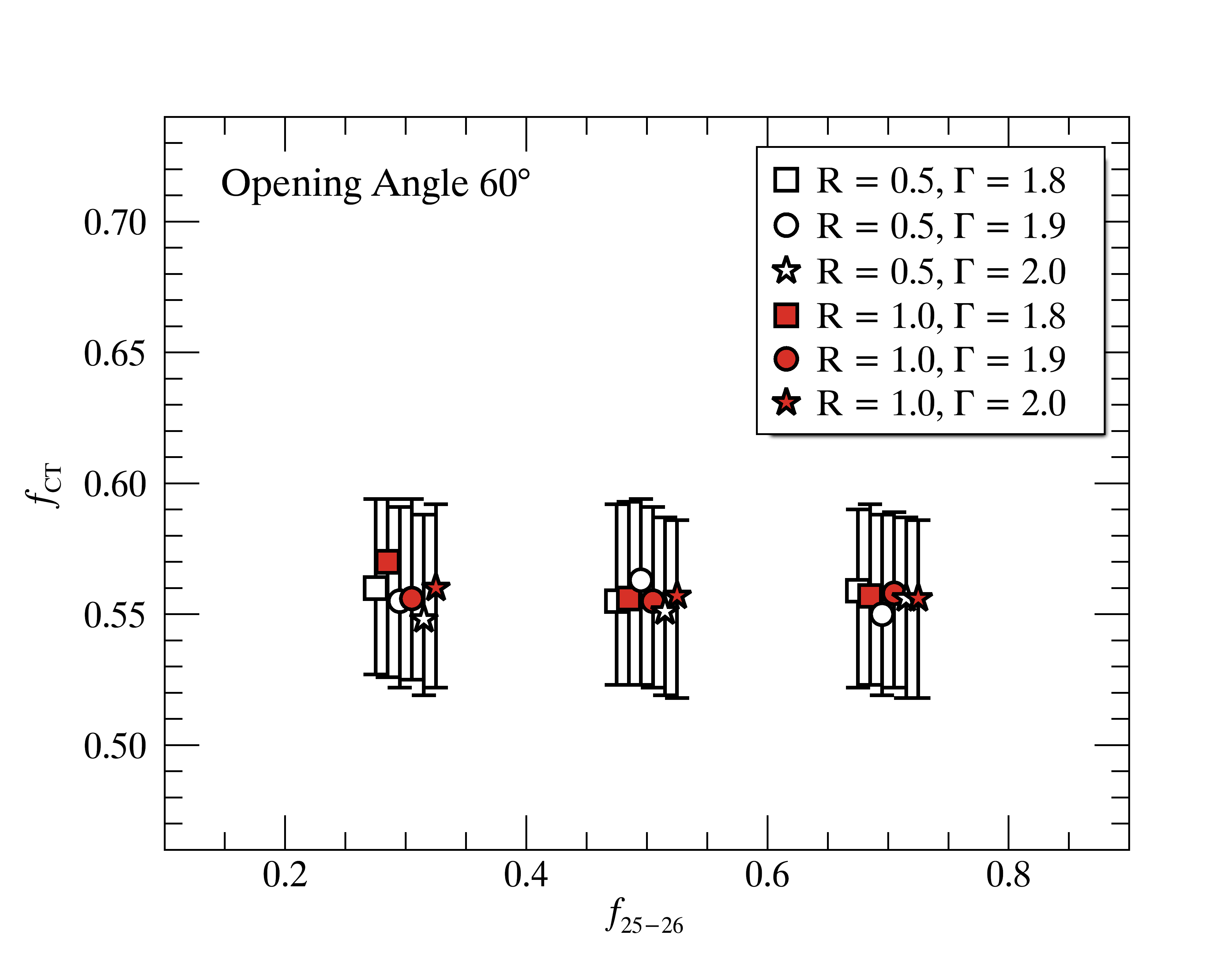}
    \caption{Additional parameter tests---$f_{\text{CT}}$. Left: Compton-thick fraction $f_{\text{CT}}$ as a function of $f_{\text{25--26}}$ \{0.3,\,0.5\,,0.7\} for an opening angle of \mbox{$\theta_{\text{OA}}$ = 30$\degr$}. Variations on reflection strength $R$ \{0.5,\,1.0\} (open and closed symbols) and photon index $\Gamma$ \{1.8,\,1.9,\,2.0\} (squares, circles, and stars) show negligible impact on $f_{\text{CT}}$. Symbols are plotted with horizontal offset for visual clarity. Right: Similar plot, but for an opening angle of \mbox{$\theta_{\text{OA}}$ = 60$\degr$}.}
    \label{fig:params_fct}
\end{figure}
\vspace{-7mm}
\begin{figure}[h]
    \epsscale{0.8}
    \plottwo{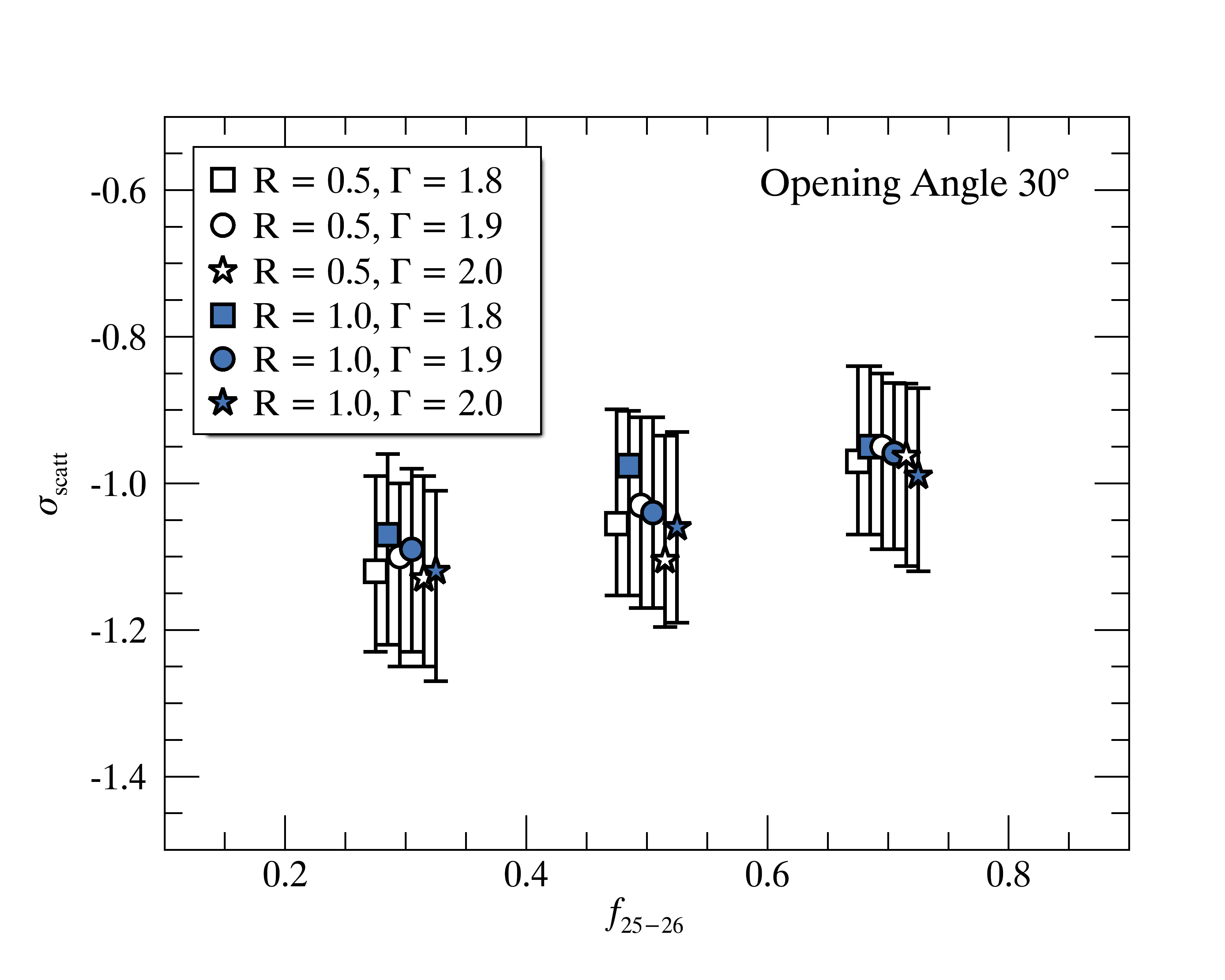}{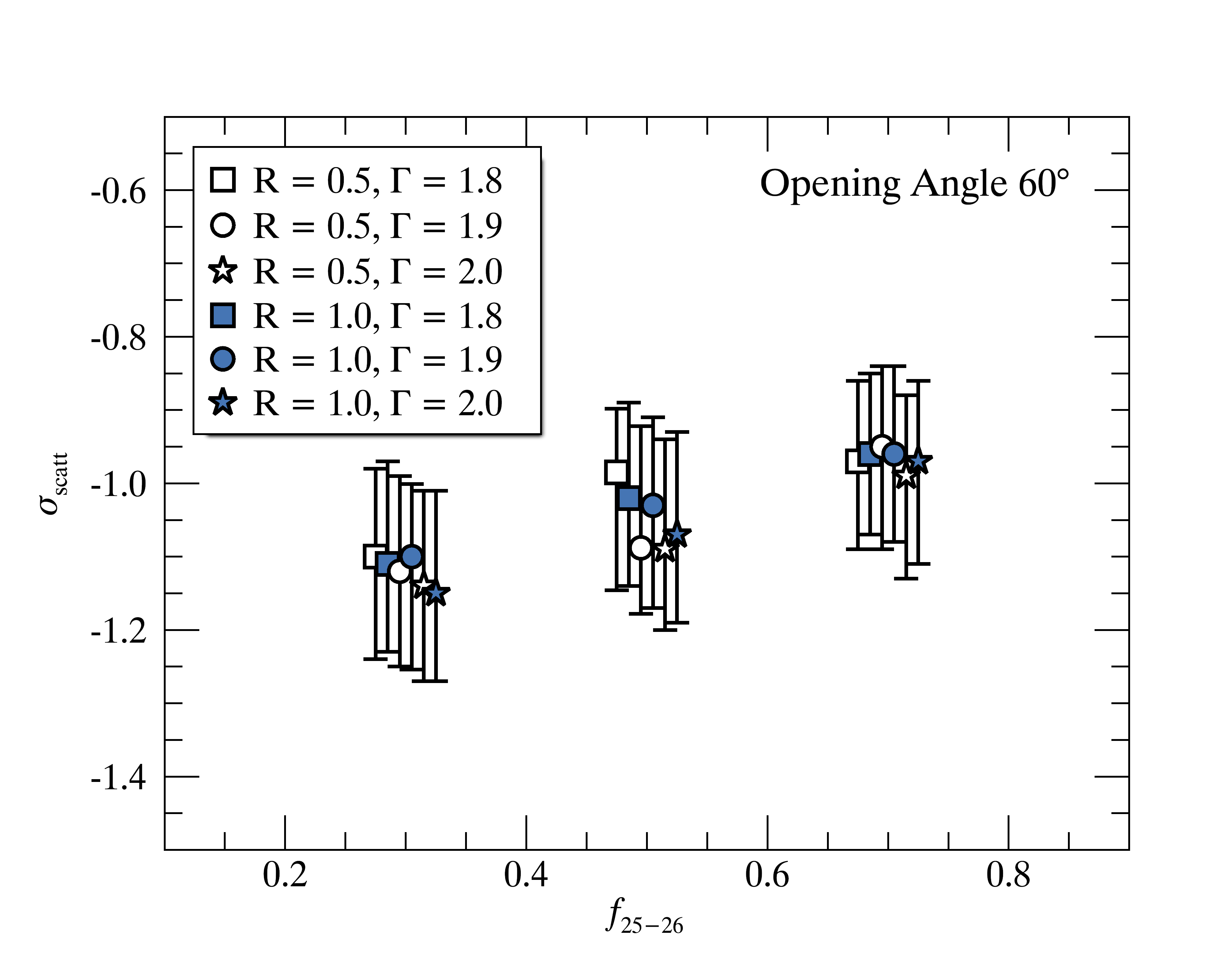}
    \caption{Additional parameter tests---$f_{\text{scatt}}$. Left: Uncertainties on scattering fraction $\sigma_{\text{scatt}}$ as a function of $f_{\text{25--26}}$ \{0.3,\,0.5\,,0.7\} for an opening angle of $\theta_{\text{OA}}$ = 30$\degr$. Variations on reflection strength $R$ \{0.5,\,1.0\} (open and closed symbols) and photon index $\Gamma$ \{1.8,\,1.9,\,2.0\} (squares, circles, and stars) show negligible impact on $\sigma_{\text{scatt}}$. Symbols are plotted with horizontal offsets for visual clarity. Right: Similar plot, but for an opening angle of \mbox{$\theta_{\text{OA}}$ = 60$\degr$}.}
    \label{fig:params_fscat}
\end{figure}

%%%%%%%%%%%%%%%%%%%%%%%%%%%%%%%%%%%%%%%%%%%%%%%%%%%%%%%%%%%
%%      Appendix: Log-Likelihood Function
%%%%%%%%%%%%%%%%%%%%%%%%%%%%%%%%%%%%%%%%%%%%%%%%%%%%%%%%%%%
\clearpage
\restartappendixnumbering
\section{Log-Likelihood Function}\label{sec:app_likelihood}
\begin{figure}[h]
    \epsscale{0.4}
    \plotone{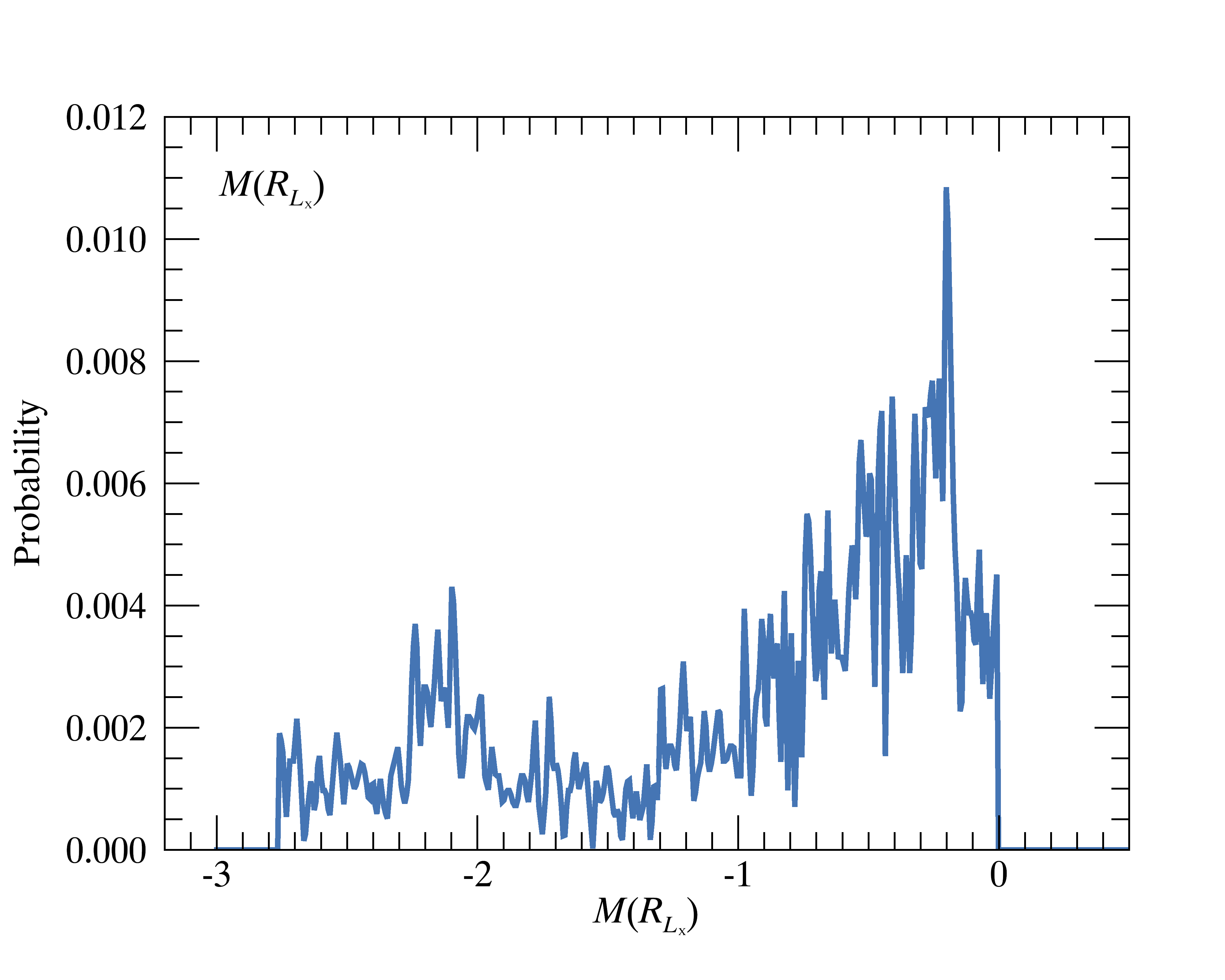}
    \caption{Example $M(R_{L_{\text{X}}})$ given a set of parameters \{$f_{\text{CT}}$,\,$f_{\text{scatt}}$\}.} 
    \label{fig:mrlx}
\end{figure}
\begin{figure}[h]
    \epsscale{0.8}
    \plottwo{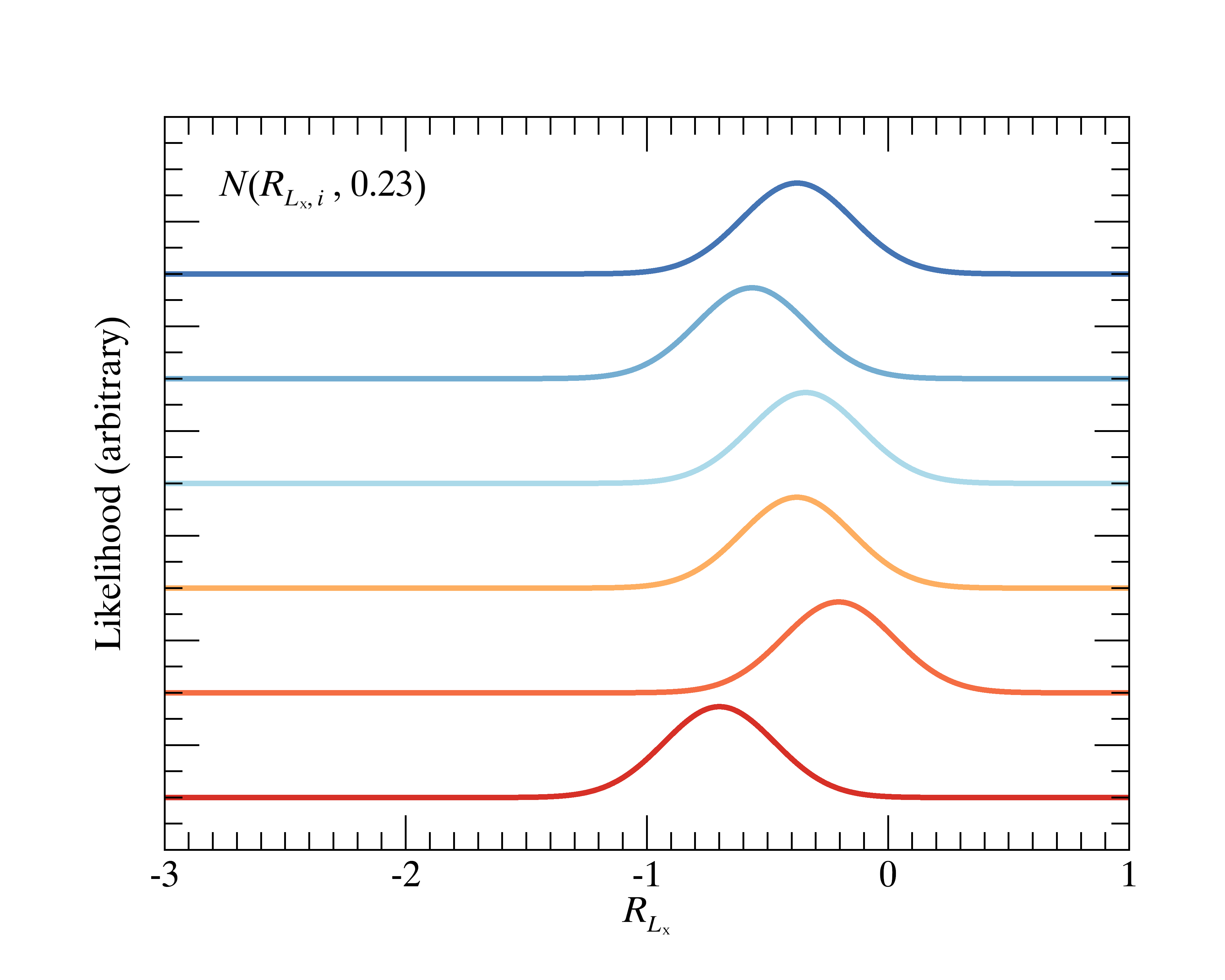}{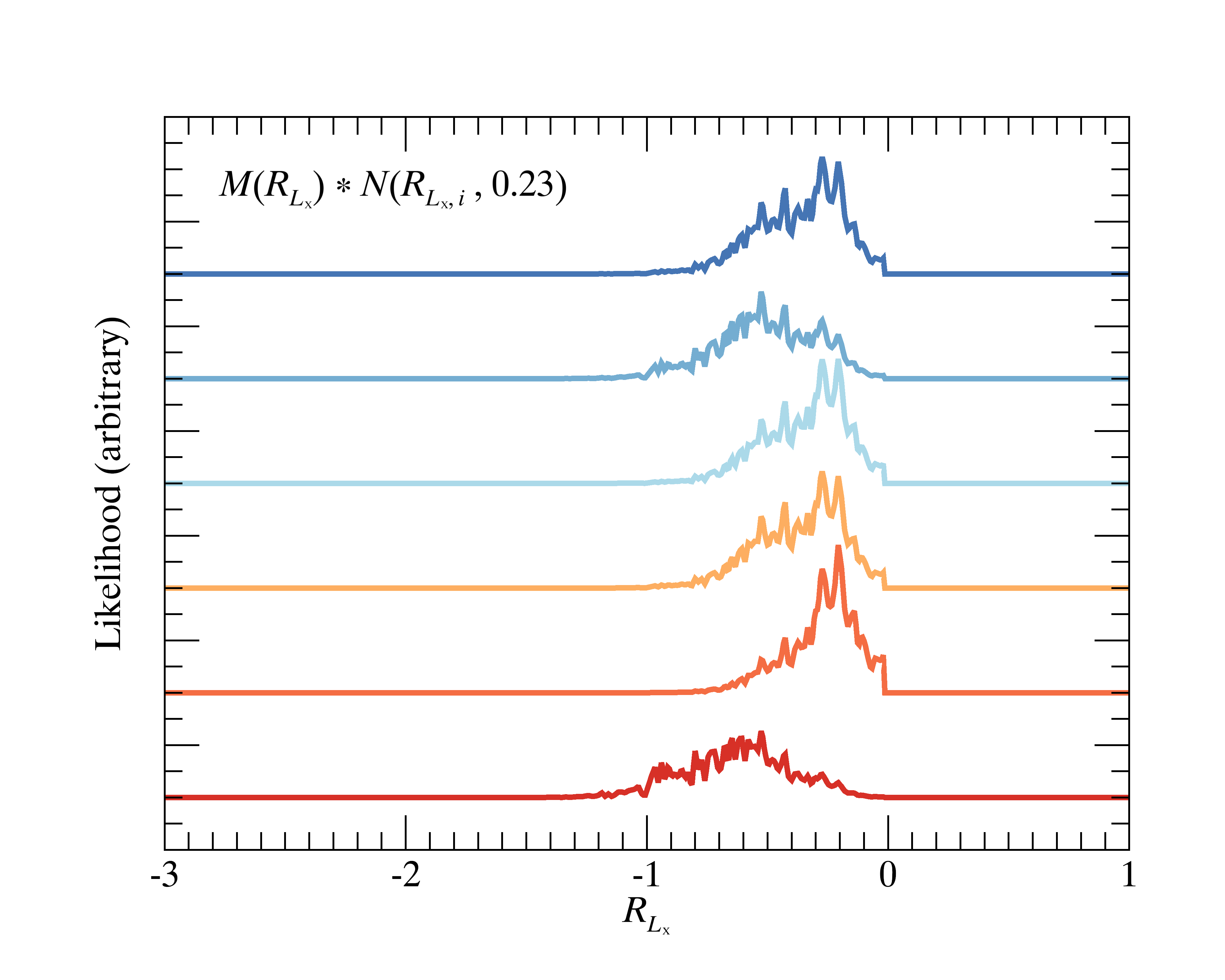}
    \caption{Integrand of the likelihood function---first term. Left: Observed $(R_{L_{\text{X}},i},\,\text{0.23})$ distribution. Right: Observed $(R_{L_{\text{X}},i},\,\text{0.23})$ distribution convolved with the model $M(R_{L_{\text{X}}})$. Note that each curve represents an X-ray detected source from our sample. The first term of the likelihood function (Equation \ref{eq:likelihood}) is the sum of the logarithm of the integral of these functions.} 
    \label{fig:likelihood_first}
\end{figure}
\begin{figure}[h]
    \epsscale{0.8}
    \plottwo{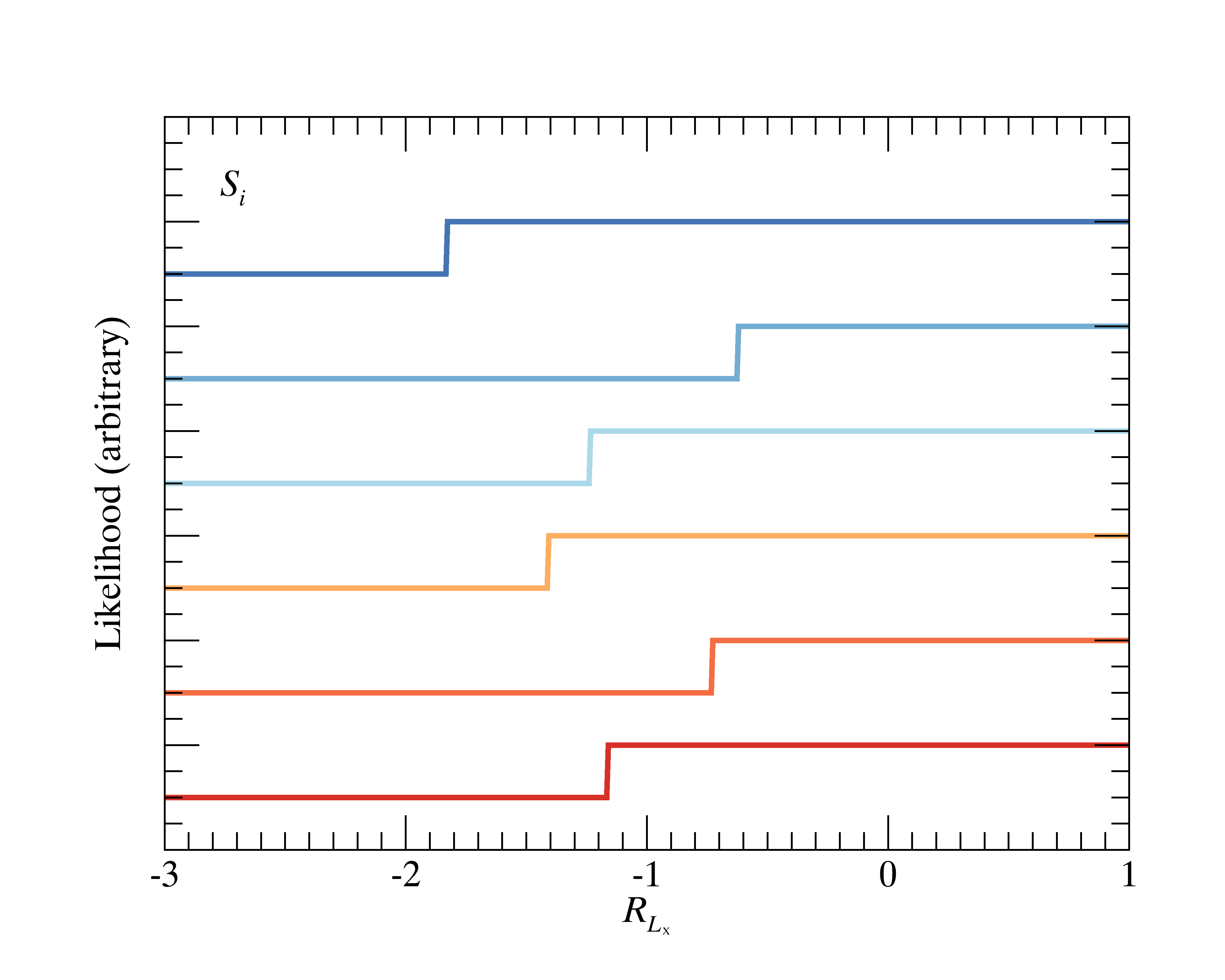}{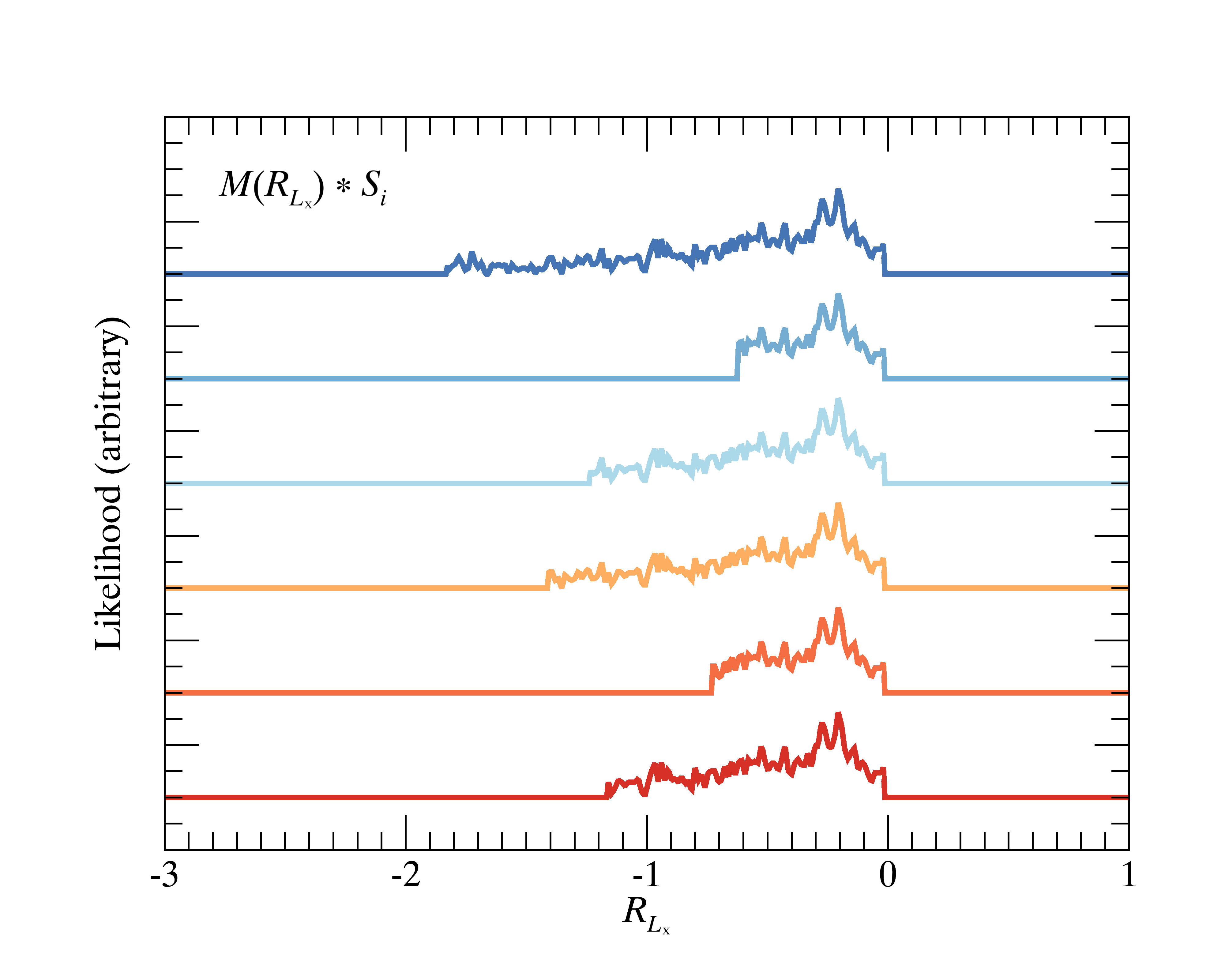}
    \caption{Integrand of the likelihood function---second term. Left: Sensitivity function $S_i(R_{L_{\text{X-lim}}})$. Right: Sensitivity function $S_i(R_{L_{\text{X-lim}}})$ convolved with the model $M(R_{L_{\text{X}}})$. Note that each curve represents a different object in the sample (both X-ray detected and non-detected). The second term of the likelihood function (Equation \ref{eq:likelihood}) is the sum of these functions, integrated over $R_{L_{\text{X}}}$.} 
    \label{fig:likelihood_second}
\end{figure}

%%%%%%%%%%%%%%%%%%%%%%%%%%%%%%%%%%%%%%%%%%%%%%%%%%%%%%%%%%%
%%
%%      Bibliography
%%
%%%%%%%%%%%%%%%%%%%%%%%%%%%%%%%%%%%%%%%%%%%%%%%%%%%%%%%%%%%
\clearpage
\bibliography{ccarroll}{}
\bibliographystyle{aasjournal}

\end{document}